\documentclass[article,preprint,preprintnumbers,amsmath,amssymb,nofootinbib,aps]{revtex4}
\pdfoutput=1
\usepackage{epsfig}

\usepackage{tikz}
\usepackage{graphicx}
\usepackage{dcolumn}
\usepackage{bm}
\usepackage{amsmath}
\usepackage{amssymb}
\usepackage{latexsym}
\usepackage{color}
\usepackage{changes}
\usepackage{mathrsfs}
\usepackage{float}
\usepackage{soul}
\usetikzlibrary{shapes,snakes}
\usepackage{verbatim}

\newcommand{\be}{\begin{equation}}
\newcommand{\ee}{\end{equation}}
\newcommand{\bea}{\begin{eqnarray}}
\newcommand{\eea}{\end{eqnarray}}

\begin{document}


\title{Direct detections of the Axion-like particle Revisited}

\author{Wei Chao}
\email{chaowei@bnu.edu.cn}
\affiliation{
	Center for Advanced Quantum Studies, Department of Physics,
	Beijing Normal University, Beijing 100875, China
}
\author{Jing-Jing Feng}
\email{fengjj@mail.bnu.edu.cn}
\affiliation{
	Center for Advanced Quantum Studies, Department of Physics,
	Beijing Normal University, Beijing 100875, China
}
\author{Ming-Jie Jin}
\email{jinmj@bnu.edu.cn}
\affiliation{
	Center for Advanced Quantum Studies, Department of Physics,
	Beijing Normal University, Beijing 100875, China
}

\def\zt{\mbox{$z_t$}}

\begin{abstract}
Axion-like particles (ALPs) are promising dark matter candidates. Their signals in direct detection experiments arise from the well-known inverse Primakoff effect or the inverse Compton scattering of the ALPs with the electron. In this paper, we revisit the direct detection of ALP by carefully considering the interference between the inverse Primakoff amplitude and the inverse Compton amplitude  in the scattering process $a+e \to e+\gamma$ for the first time. It shows that the contribution of the interference term turns to be dominated in the scattering for a large ALP energy. Given the new analytical formula, signals or constraints of ALP couplings in various projected experiments are investigated. Our results show that these experiments may put strong constraints on ALP couplings for relatively heavy ALP. We further study projected constraints on the ALP from the JUNO experiment, which shows competitive constraints on ALP couplings using a ten-year exposure.

\end{abstract}

\maketitle



\section{Introduction} 

The strong CP problem finds a solution via the  Peccei-Quinn mechanism~\cite{Peccei:1977hh,Peccei:1977ur}, which gives rise to a fascinating prediction: the existence of a new particle called the axion, as outlined in Refs~\cite{Weinberg:1977ma,Wilczek:1977pj}. These ultra-light, weakly interacting particles, such as axions or axion-like particles (ALPs), have also gained prominence as potential dark matter (DM) candidates within the diverse theories of DM~\cite{Kim:2008hd,Feng:2010gw,Tam:2011tz,Marsh:2015xka}. Since they can couple to the Standard Model (SM) particles such as photon and electron, extensive experimental works~\cite{Spector:2019ooq,CAST:2017uph,LUX:2017glr,XENON:2022ltv} have been devoted to searching for such particles through the axion-photon coupling or the axion-electron coupling, and stringent limits on these couplings have already been imposed.

The Sun is an excellent source of axions, offering an easily obtainable flux of axions generated in the center of the Sun via various processes~\cite{Redondo:2013wwa,Derbin:2012yk}. However, the endeavor to detect axions or ALPs is an arduous undertaking, primarily due to their extremely weak interactions with SM particles. Fortunately, Sikivie~\cite{Sikivie:1983ip} made a breakthrough by showing that these elusive ALPs may be detected when they convert into photons in the magnetic fields in 1983. This discovery presented a viable method for the detection of ALPs. Currently, many experiments are carried out searching for ALPs, of which the constraint given by the CERN Axion Solar Telescope (CAST) is $g_{a\gamma}<6.6\times10^{-11}(\rm GeV^{-1})$~\cite{CAST:2007jps,CAST:2017uph}, with $g_{a\gamma}$ the coupling of the ALP to diphoton. Still, more stringent bounds arise from  astrophysical observations, of which the stellar evolution, determined using the R2 parameter,  the ratio of stellar populations on the asymptotic giant branch to Horizontal Branch stars in globular clusters, gives $ g_{a\gamma}<4.7\times10^{-11}(\rm GeV^{-1})$~\cite{Dolan:2022kul}. 

In addition to searching for ALPs with a Helioscope, there is currently a significant focus on utilizing DM direct detection experiments based on underground laboratories\cite{PandaX:2017ock,Knapen:2016moh,ATLAS:2020hii,DarkSide:2022knj} to either directly detect ALPs or set novel constraints on ALP couplings in projected direct detection experiments. Either solar axions or heavy axion DM with large kinetic energy~\cite{Arisaka:2012pb,Irastorza:2018dyq} can be detected in these experiments. Assuming that ALPs couple to neutrinos, the supernovae neutrinos can also boost the galactic DM ALPs via elastic scattering, resulting in relativistic ALPs~\cite{Carenza:2022som}. In this paper, we will consider the direct detection signals of the ALP arising from the following three sources: (1) supernova neutrino-boosted ALPs ($ v\sim c $); (2) solar axion; (3) galactic DM ALP ($ v\sim10^{-3}c $) with large kinetic energy.

In this paper, we focus on the scattering process $a+ e\to e + \gamma$, where $a$ is an ALP, $e$ and $\gamma$ are electron and photon, respectively. This process, if mediated by the dimension-five operator $g_{a\gamma}a F_{\mu\nu}\tilde{F}_{\mu\nu}$, is called inverse Primakoff~(IP) scattering~\cite{Buchmuller:1989rb,Creswick:1997pg}. Alternatively, the scattering mediated by the dimension-four operator $g_{ae} a \bar{e}i \gamma^5 e$ is called the inverse Compton~(IC) scattering~\cite{Avignone:1988bv,Chanda:1987ax,Brodsky:1986mi}. As shown in Fig.~\ref{feynman} (a), (b) and (c), the coherent enhancement factor for the inverse Primakoff scattering is $Z^2$, with $Z$ the atomic number of the target nucleus, which is similar to the Coulomb scattering process, while the coherent enhancement factor for the inverse Compton scattering is $Z$. Usually these two processes are separately considered in the literature by setting either $g_{a\gamma}^{}$ or $g_{ae}^{}$ to be zero. However, the axion-photon and the axion-electron couplings may exist simultaneously in a real case, so one needs to account for the contribution to the scattering induced by the interference of two interactions, which is done in this paper. Our calculation shows that this contribution can be dominant for a large axion energy $E_a$, as can be seen in Fig.~\ref{fig2}. The complete calculation helps to improve constraints on ALP couplings in projected DM direct detection experiments such as XENON, PandaX, etc. 

Apart from the DM direct detection experiments mentioned above, an increasing number of researchers are interested in probing axions or ALPs using neutrino detectors, such as Super-K\cite{Li:2022xqh}, JUNO\cite{Lucente:2022esm}, and others\cite{Dent:2019ueq,Bhusal:2020bvx}. The strength of neutrino experiments lies in their substantial fiducial mass ($ \sim $20kton), which helps to improve the exclusion limits. However, their drawback is that they can only search for ALPs with high energies due to the threshold of the detector. In this paper, we focus on the detectability of the supernova neutrino-boosted ALP, whose energy can reach MeV, using the JUNO detector. Projected constraints on the ALP coupling at the $ 90\% $ confidence level~(C.L.) is presented in the Fig.~\ref{fig6}.

The remaining of the paper is organized as follows: in Section~\ref{sec: IPICIT}, we calculate the scattering cross section $a+ e\to e + \gamma$ induced the inverse Primakoff process, the inverse Compton process and the interference between these two terms. We discuss three fluxes of incoming axions: supernovae neutrinos boosted axions, solar axions, and DM axions in Section~\ref{sec: direct detection}. Then, we study the limitations of the axion couplings $g_{a\gamma}$ and $g_{ae}$ in DM direct detection experiments. The constraints of JUNO are given in Section~\ref{sec:  JUNO}. Finally, summary remarks are given in the Section~\ref{sec: conclusion}.

\section{\label{sec: IPICIT} ALP-electron Scattering}
ALPs have gained attention in particle physics as they may play an important role in addressing various problems in terrestrial experiments and astrophysical observations, including DM and the strong CP problem. ALPs are assumed to be super-light and weakly interacting particles, and are predicted in certain extensions to the SM with spontaneous breaking global symmetries. Despite the possibility for ALPs to interact with a wide range of particles in various forms, most ALP models actually propose interactions specifically with photons and electrons with Lagrangian of the following form: 
\begin{eqnarray}
	\mathcal{L}=-\frac{1}{4} g_{a\gamma} a F^{\mu\nu} \tilde{F}_{\mu\nu}+ g_{ae} a \bar{e}i \gamma^5 e.
\end{eqnarray}
where $ a $ stands for the ALP. $ F^{\mu \nu} $ is the electromagnetic field strength and its dual  $\tilde{F}_{\mu\nu}=\dfrac{1}{2} \epsilon_{\mu \nu \alpha \beta} F^{\alpha \beta} $. $ g_{a\gamma} $ and $ g_{ae} $ denote the axion-photon and axion-electron couplings, respectively. 
For the detection of ALPs, they can be probed  via oscillations of ALP into the photon in a Helioscope with constraint put on the coupling $g_{a\gamma}$.
They can also be detected by the Axio-electric effect by measuring the rate of atomic ionization induced by the absorption of ALP of the energies up to 100 keV~\cite{Derevianko:2010kz}.  
Alternatively, they can be detected in direct detection experiments via the inverse Compton scattering~\cite{Donnelly:1978ty,Borexino:2008wiu,Dent:2019ueq} or the inverse Primakoff process~\cite{SOLAX:1997lpz,Abe:2020mcs,Buchmuller:1989rb,Dent:2020jhf,Gao:2020wer} with the photon as the signal.  
In this paper, we focus on the photon signal of ALPs in underground detectors, so we pay special attention to the  inverse Compton scattering and the inverse Primakoff process, as well as their interference effect assuming $g_{ae}$ and $g_{a\gamma}$ exist simultaneously. 

\begin{figure}[t]
	\begin{center}
		\includegraphics[height=4cm,width=16cm]{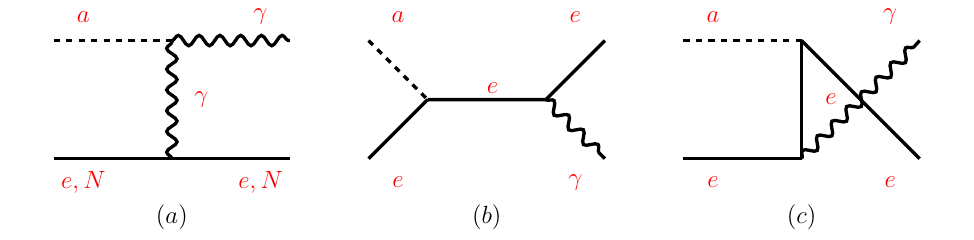}
		\caption{Tree-level Feynman diagrams illustrating (a) the inverse Primakoff  process  and (b), (c) the inverse Compton process. }\label{feynman}
	\end{center}
\end{figure}

Many experiments employ the inverse Primakoff scattering process of the ALP with atoms as the detection channel, where the incoming ALPs are converted into photons in the electromagnetic field of the atom. It utilizes a screened Coulomb potential to portray the electrostatic field in the target atom. This potential takes the form of the Yukawa potential, with the screening length denoted as $r_0$. For the inverse Compton scattering process as shown in the Fig.~\ref{feynman} (b) and (c), the collision of the ALP with the electron in the detector results in a photon in the final state. It should be mentioned that there is IP-like scattering process that also contribute to the $a+ e\to e+ \gamma$ scattering, as shown in the Fig.~\ref{feynman} (a), the interference of which with the traditional inverse Compton process hasn't been explicitly calculated.

The differential scattering cross section of the IP process is directly governed by the parameters $ g_{a\gamma} $ and $ m_a $, which determine the strength and characteristics of the interaction. The cross section of this process is obtained as
\begin{eqnarray}
		\frac{d\sigma^{\rm{IP}}}{d\Omega}=\frac{\alpha}{8 \pi}g^2_{a \gamma}\left[\frac{1}{(E_a+m_{e,N}-p_a\cos{\theta})}\dfrac{E^2_\gamma}{p_a^3}(E_a-E_\gamma+2m_{e,N}) \right] F^2_a(\boldsymbol{q}^2) \sin^2{\theta},
        ~\label{IP}
\end{eqnarray}
where $ \alpha=e^2/4 \pi $ is the fine structure constant and $ m_{e, N} $ is the electron or nucleus mass. For the incoming ALP of the mass $ m_a $, its energy and momentum are $ E_a $ and $ \boldsymbol{p_a} $, respectively, $p_a=|\boldsymbol{p_a}|$. The energy and momentum of the outgoing photon are $ E_\gamma $ and $ \boldsymbol{p_\gamma} $. $ \theta $ is the scattering angle and $ \boldsymbol{q}=\boldsymbol{p_\gamma}-\boldsymbol{p_a} $ is the momentum transfer. An alternative method utilized in this paper to approximate the form factor $F_a(\boldsymbol{q}^2)$ involves assuming a screened Coulomb potential arising from the electrostatic field, resulting in $F_a(\boldsymbol{q}^2)=\frac{Z {p}_a^2}{\boldsymbol{q}^2+r^{-2}_0}$~\cite{SOLAX:1997lpz,Abe:2020mcs}. According to the energy conservation, one has
\begin{eqnarray}
		E_\gamma=\frac{2m_{e, N} E_a+m^2_a}{2(E_a+m_{e,N}-{p_a}\cos{\theta})}.
         ~\label{outgoing photon energy}
\end{eqnarray}
Taking  Eq.(\ref{outgoing photon energy}) into  Eq.(\ref{IP}) and then integrate over the solid angle, we can obtain $ \sigma^{\rm{IP}}(g_{a \gamma},E_a,m_a) $.

In the traditional calculation of the IP process, the nucleus recoil is neglected resulting in $E_\gamma=E_a$. 
Furthermore, if we ignore the mass of the axion, the transfer momentum can be simplified as $|\boldsymbol{q}|=2p_a\sin{(\theta/2)}$ and the factor $(E_a+m_{e,N}-p_a\cos{\theta})$ can be approximated as $m_{e, N}$. 
In this way, our result in Eq.~(\ref{IP}) can be simplified to the differential cross section which appear in most of the literature~\cite{SOLAX:1997lpz,Abe:2020mcs,Buchmuller:1989rb,Dent:2020jhf,Gao:2020wer},
\begin{eqnarray}
	\begin{aligned}
		\frac{d\sigma_0^{\rm{IP}}}{d\Omega}
		&=\frac{\alpha}{4 \pi}g^2_{a \gamma}F^2_a(\boldsymbol{q}^2)\sin^2{\theta}
		=\frac{\alpha}{16 \pi}g^2_{a \gamma}\frac{\boldsymbol{q}^2}{{p}_a^2}\left(4-\frac{\boldsymbol{q}^2}{{p}_a^2}\right)F^2_a(\boldsymbol{q}^2)  \; .
	\end{aligned}\label{IP0}
\end{eqnarray}

The differential cross section for the inverse Compton scattering is given as~\cite{Donnelly:1978ty,Borexino:2008wiu,Dent:2019ueq}
\begin{eqnarray}
		\frac{d\sigma^{\rm{IC}}}{d\Omega}=\frac{Z g^2_{ae}}{8\pi m^2_e}\frac{\alpha E_\gamma}{{p}_a}\left(1+\frac{4m^2_e E^2_\gamma}{(2m_e E_a+m^2_a)^2}-\frac{4m_e E_\gamma}{(2m_e E_a+m^2_a)}-\frac{4m^2_a{p}_a^2m_eE_\gamma\sin^2{\theta}}{(2m_e E_a+m^2_a)^3}\right) \; , 
	    ~\label{IC}
\end{eqnarray}
which is enhanced by the  atomic number  $Z$. 

Most importantly, if there are axion-photon and axion-electron couplings simultaneously (i.e., both $ g_{a \gamma} $ and $ g_{ae} $ are non-zero), the total squared matrix elements of the combined process exhibit interference between the traditional inverse Compton scattering terms given in Fig.~\ref{feynman} (b) and (c) and the IP-like term given in Fig.~\ref{feynman} (a). 
To assess the contributions of the interference terms, we define the cross section arising from the interference terms (IT) as $\sigma^{\rm IT}$, which can be written as
\begin{eqnarray}
		\frac{d\sigma^{\rm{IT}}}{d\Omega}=\frac{\alpha}{8\pi}\frac{Z g_{a \gamma}g_{ae} E^2_\gamma}{(m_e+E_a-{p}_a\cos{\theta})}\frac{{p}_a\sin^2{\theta}}{(\boldsymbol{q}^2+r^{-2}_0)}\left(\frac{1}{m_e+E_a-{p}_a\cos{\theta}}+\frac{1}{m_e}\right).
		~\label{IT}
\end{eqnarray}
It should be mentioned that $\sigma^{\rm IT}$ is only enhanced by the factor $Z$, which is because we only account the interference of the inverse Compton scattering and the IP-like process induced by the Coulomb potential of the single electron. 

\begin{figure}[t]
	\begin{center}
		\includegraphics[height=6cm,width=9cm]{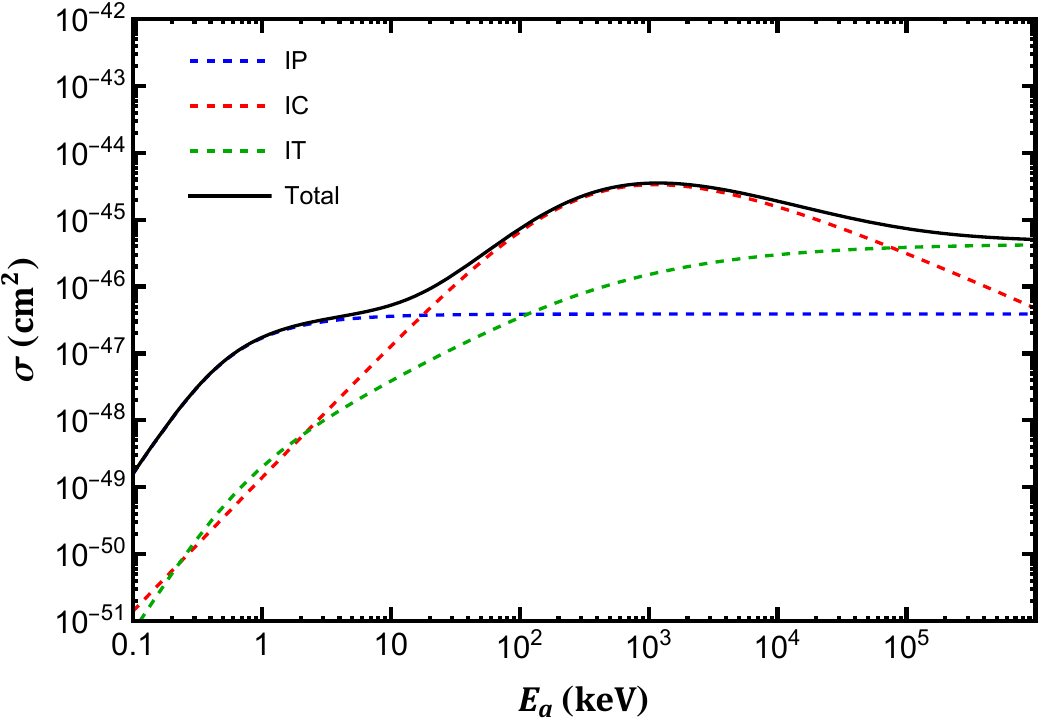}
		\caption{The cross sections inverse Primakoff~(blue dashed line), inverse Compton(red dashed line) and interference terms(green dashed line) conversion of axion into photon. The black solid line represents the total cross section. } \label{fig2}
	\end{center}
\end{figure}

Finally, the total cross section is $\sigma^{\rm Tot}=\sigma^{\rm{IP}}+\sigma^{\rm{IC}}+\sigma^{\rm{IT}}$. We explore the contribution of each term, as depicted in Fig.~\ref{fig2}, varies with the ALP energy ($ E_a $). The depicted figure represents a scenario with parameters set as $g_{a\gamma}=10^{-10}~{(\rm{GeV}^{-1})}$, $ g_{ae}=10^{-11}$, $ Z=54~({\rm Xe})$, $r_0=2.45\mathring{{\rm{A}}}~({\rm LXe})$, $m_a=10^{-9}~{\rm{keV}}$. If we alter the coupling constants as well as the target material, the contributions of each process will undergo changes.

\section{\label{sec: direct detection} Direct Detection Signal}

In this section, we discuss the photon signal of ALP in DM direct detection experiments by accounting the contribution of the interference given in the last section. We will first discuss the ALP flux used in the calculation, then we show constraints of current and projected direct detection experiments on ALP couplings. 

\subsection{ALP Flux}

\subsubsection{neutrino-boosted ALP flux }

The possibility of boosting ALPs by elastic scattering of supernova neutrinos with ALPs is discussed in Refs~\cite{Carenza:2022som,Lin:2022dbl}. Consequently, the ALP energy experiences a substantial amplification, leading them to exhibit characteristics of relativistic particles. Although the scattering cross section is small, the huge neutrino flux greatly enhances the ALP flux, which helps to increase the number event of the direct detection. We conservatively estimate the ALP flux by ignoring the effect of neutrino oscillations, and the boosted ALP flux given in the Ref.~\cite{Carenza:2022som} is applied in our analysis.

\subsubsection{solar axion flux}
Both axion-photon and axion-electron couplings are important parameters to determine the solar axion flux. The axion-electron coupling induces a large number of reactions that are significant. The most important processes are the ABC reactions: Atomic axio-recombination\cite{Dimopoulos:1986mi,Pospelov:2008jk} and Atomic axio-deexcitation, axio-Bremsstrahlung in electron-Ion\cite{Zhitnitsky:1979cn,Krauss:1984gm} or electron-electron collisions, Compton scattering\cite{Mikaelian:1978jg,Fukugita:1982ep,Fukugita:1982gn}. The axion flux, arising from the Primakoff effect\cite{Raffelt:1985nk,Raffelt:1987np} induced by axion-photon coupling, cannot be disregarded in certain cases. In this paper, the ALP flux comprises both the ABC process and the Primakoff process. We reference the solar axion flux reported in Ref.~\cite{Redondo:2013wwa}(for details regarding the solar axion flux, see Ref.~\cite{Redondo:2013wwa,Derbin:2012yk}). 
\begin{figure}[t]
	\begin{center}
		\includegraphics[height=6cm,width=8cm]{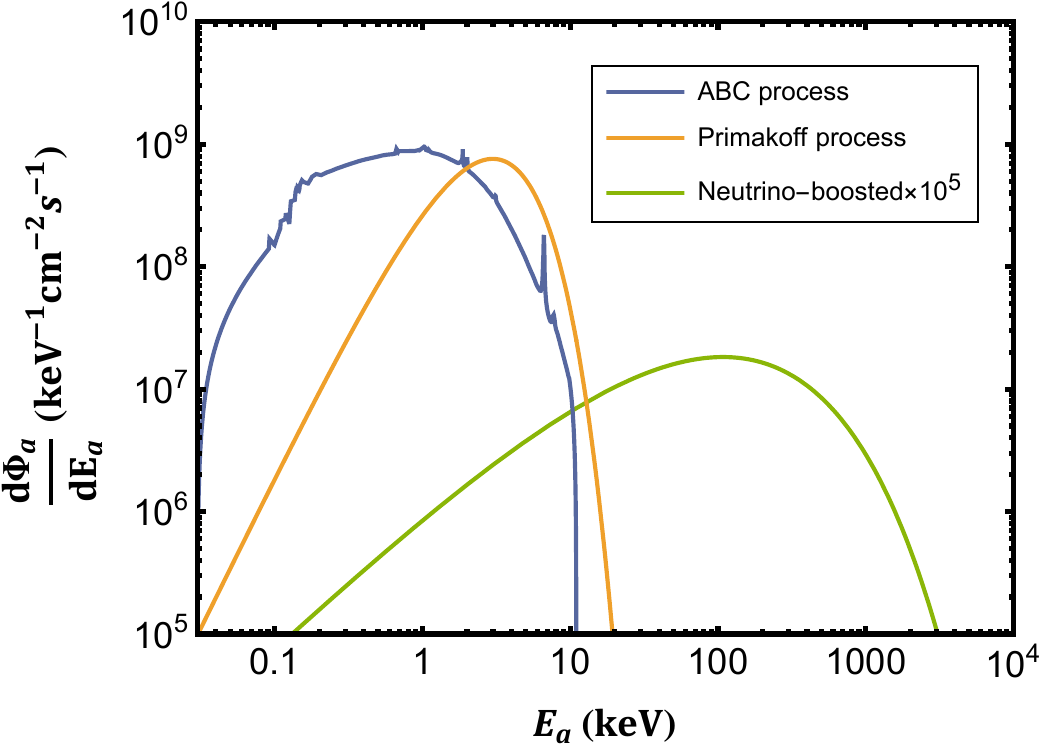}
		\caption{The blue line represents the solar axion produced by the ABC processes using $g_{ae}=10^{-13}$. The orange line represents the solar axion generated by the Primakoff process using $g_{a\gamma}=10^{-11}(\rm GeV^{-1})$. Neutrino-boosted ALP flux is shown as green line with $m_a=10^{-15}\rm keV$. Note that has been scaled up by a factor $10^5$ to make it visible.} \label{flux}
	\end{center}
\end{figure}

\subsubsection{DM ALP flux}
If we assume that axions exclusively make up the entirety of the galactic DM density, the total flux of DM ALP is
\begin{eqnarray}
		\Phi_a=\frac{\rho_{\rm{DM}}}{m_a}v_a.
\end{eqnarray}
Where $m_a$,~$v_a$ and $\rho_a$ are the mass, velocity and the local energy density of the ALP, respectively. Precisely, the local DM density in the solar system is taken as 
$\rho_{\rm{DM}}=(0.2-0.6)\rm{GeV}/cm^3$~\cite{Planck:2018vyg,ParticleDataGroup:2022pth}. The flux of DM ALP can be expressed by 
\begin{eqnarray}
	\begin{aligned}
		\Phi_a&= {1 {\rm keV} \over m_a }\times 9\times10^{12} ~~(\rm cm^{-2} \cdot s^{-1}).
	\end{aligned}
	\label{dmflux}
\end{eqnarray}
by taking $ \rho_{\rm{DM}}=0.3 \rm{GeV}/cm^3 $~\cite{Arisaka:2012pb,Branca:2016rez} and $v_a\sim 10^{-3}c$.  If one accounts the effect of the DM local velocity,  Eq.~(\ref{dmflux}) should be multiplied by the  factor of the Maxwell-Boltzmann distribution~\cite{Lisanti:2016jxe}. 

We show the flux of the solar ALP and neutrino boosted ALP in the Fig.~\ref{flux}, in which the blue and orange curves represent the flux induced by the ABC processes and the Primakoff process, respectively. The green curve represents the neutrino boosted ALP flux enhanced by the factor of $10^{5}$. For the DM ALP flux, we use the Eq.~(\ref{dmflux}) directly.

\subsection{Direct Detection Constraints}

In this subsection, we study signals for three types of axion fluxes in DM direct detection experiments so as to put constraints on ALP couplings. The expected number of events for signals due to the inverse Primakoff, inverse Compton and their interference in the detector is given by~\cite{Vergados:2021ejk,Lucente:2022esm}
\begin{eqnarray}
		N_{\rm{event}}=N_T \cdot \Phi_a \cdot \left(\sigma^{\rm{IP}}+\sigma^{\rm{IC}}+\sigma^{\rm{IT}}\right) \cdot \epsilon,
		\label{events}
\end{eqnarray}
where $ N_T $ is the number of targets and $ \Phi_a $ denote the flux of the incoming ALP, with the subindex $a$ denoting  three distinct types of fluxes described in the subsection A, and $ \epsilon $ represents the detector efficiency. In the following, we assume $ \epsilon=1 $ for all detection channels. 
 
We can define the signal-to-noise ratio of the signal as the quantity $ r_{\rm sn}=\frac{S}{\sqrt{N_{\rm bg}+S}} $. $ N_{\rm bg} $ is the total number of background events. The value of $ S $ is determined using the Feldman-Cousins $ 90\% $ upper confidence level of signal events, with $ r_{\rm sn} \sim 1.64 $~\cite{Lin:2022dbl}. 
Thus by comparing  the $ S $ to $ N_{\rm event} $ times the detector target exposure $ T\times W $, where $T$ and $W$ are the exposure time and the weight of the target, respectively, we can derive the $ 90\% $ C.L. limits on $ g_{ae} $ and $ g_{a \gamma} $.  For (projected) DM direct detection experiments XENONnT, PandaX-4T and PandaX-30T, the exposure are taken as 1.16, 0.63 and 30 ton-year, respectively\cite{XENON:2022ltv,PandaX-4T:2021bab,Wang:2023wrr,Liu:2017drf}.

\begin{figure}[t]
	\begin{center}
		\begin{minipage}[c]{0.5\textwidth}
			\includegraphics[height=6cm,width=8cm]{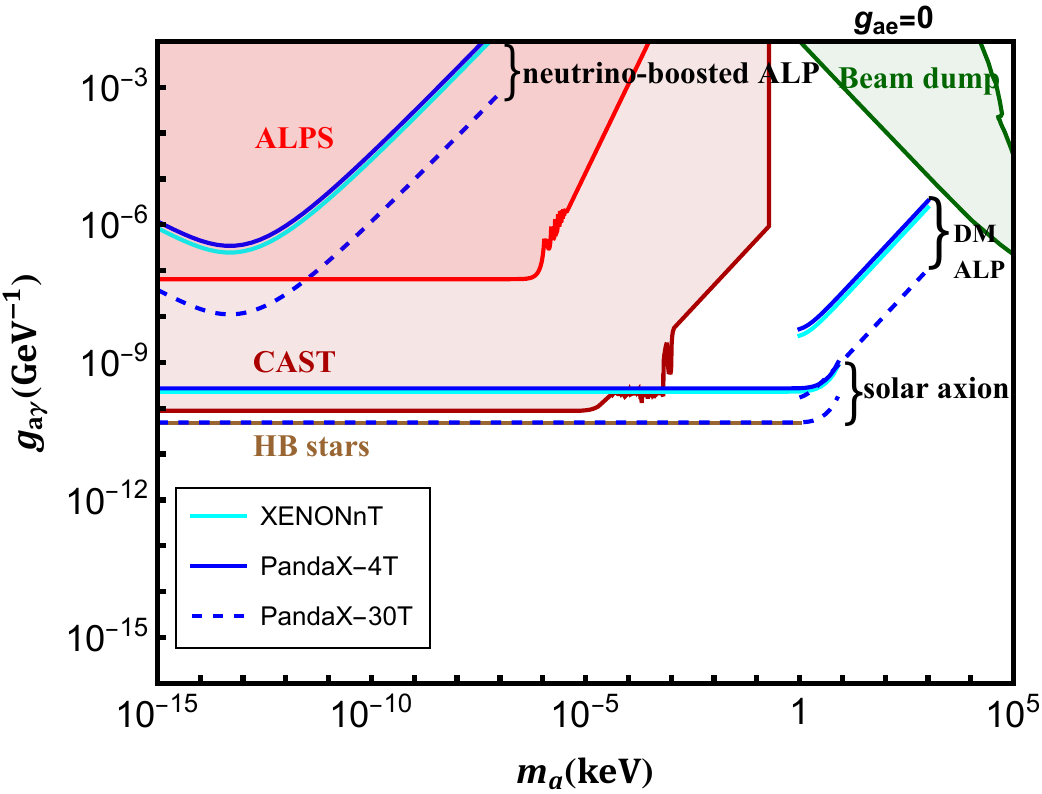}
		\end{minipage}%
		\begin{minipage}[c]{0.5\textwidth}
			\includegraphics[height=6cm,width=8cm]{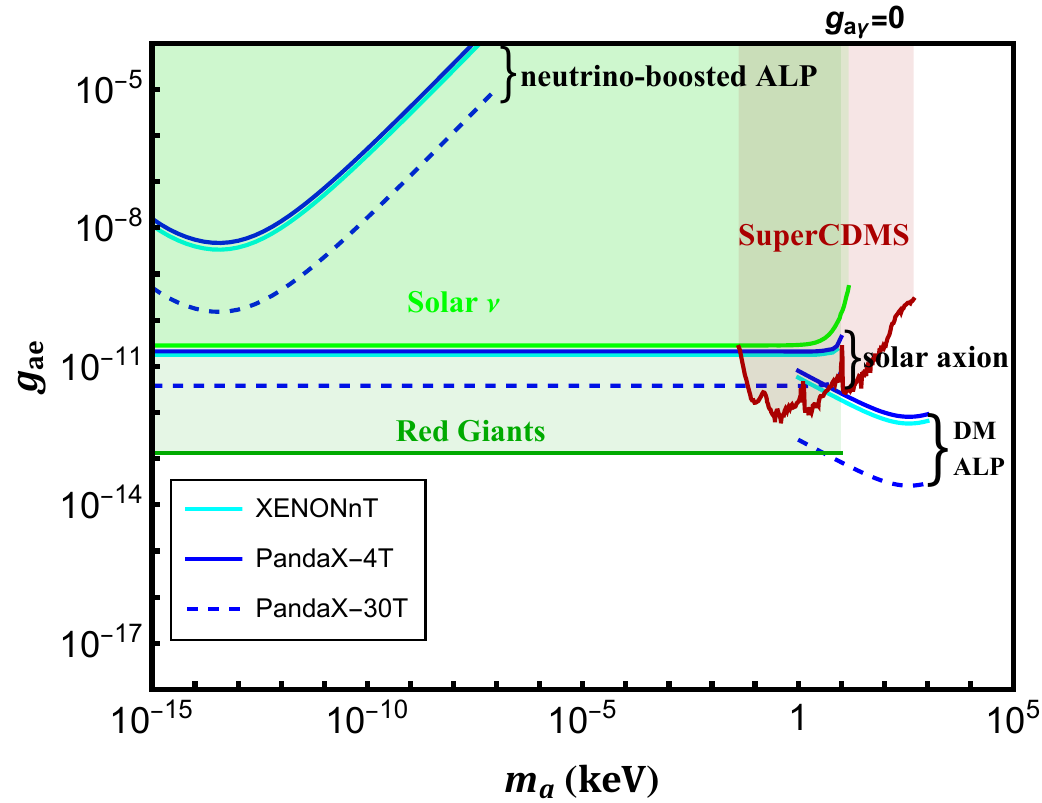}
		\end{minipage}
		\caption{$ 90\% $ C.L. limits on axion-photon(axion-electron) coupling constant versus axion mass. The blue solid lines, cyan solid lines and blue dashed lines represent the constraints of the direct detection experiments PandaX-4T, XENONnT and PandaX-30T, respectively, applied to three different initial axion fluxes: neutrino-boosted DM axion, solar axion and dark matter axion. Left-panel: Comparison of upper limit on $ g_{a \gamma} $ achieved in this analysis with other search experiments incluing ALPS~(red)~\cite{Ehret:2010mh}, CAST~(darker red)~\cite{CAST:2007jps,CAST:2017uph} and Beam dump~(green)~\cite{CHARM:1985anb,Riordan:1987aw,Dolan:2017osp,Blumlein:1990ay,NA64:2020qwq}. The constraint from Horizontal Branch stars~(brown) is also shown~\cite{Dolan:2022kul}. Right-panel: The limits shown include astrophysical bounds from solar neutrino flux~(green)~\cite{Gondolo:2008dd} and red giants~(darker green)~\cite{Capozzi:2020cbu}, underground detectors SuperCDMS~(darker red)~\cite{SuperCDMS:2019jxx}. } \label{fig3}
	\end{center}
\end{figure}

We first assume $ g_{ae}=0$ , $  g_{a\gamma}\neq0$, such that the number of expected events is simplified to $ N_{\rm{event}}=N_T \cdot \Phi_a \cdot \sigma^{\rm{IP}} $. The resulting $ 90\% $ C.L. limits are given in the left-panel of the Fig.~\ref{fig3}, which shows $ g_{a \gamma} $ versus the ALP $ m_a $ for various experiments. 
The blue solid lines, cyan solid lines and blue dashed lines represent the constraints of the direct detection experiments PandaX-4T, XENONnT and PandaX-30T, respectively. 
This plot also shows the limits from the ALPS~(red)~\cite{Ehret:2010mh}, CAST~(darker red)~\cite{CAST:2007jps,CAST:2017uph}, Beam dump~(green)~\cite{CHARM:1985anb,Riordan:1987aw,Dolan:2017osp,Blumlein:1990ay,NA64:2020qwq} and Horizontal Branch stars (brown)~\cite{Dolan:2022kul}.
Constraints on the neutrino-boosted ALP is given in the top-left corner of the Fig.~\ref{fig3} left-panel.  
Since we assume $ g_{ae}=0$, the contribution of the ABC processes to the solar ALP flux is removed,  and we only consider the ALP flux generated by the Primakoff effect, which is proportional to  $g^2_{a\gamma}$. Constraints are given as horizontal lines at the center of the  Fig.~\ref{fig3}.
While constraints induced by the DM ALP flux is given in the bottom-right corner, considering that the detector threshold for the xenon target is about ${\cal O}(1)$ keV~\cite{PandaX:2017ock,XENON:2022ltv}. 
Though these constraints don't reach the limits set by astrophysical observations, they can be improved in the future by increasing the exposure or advancing the detection technology.

\begin{figure}[t]
	\begin{center}
		\begin{minipage}[c]{0.5\textwidth}
			\includegraphics[height=6cm,width=8cm]{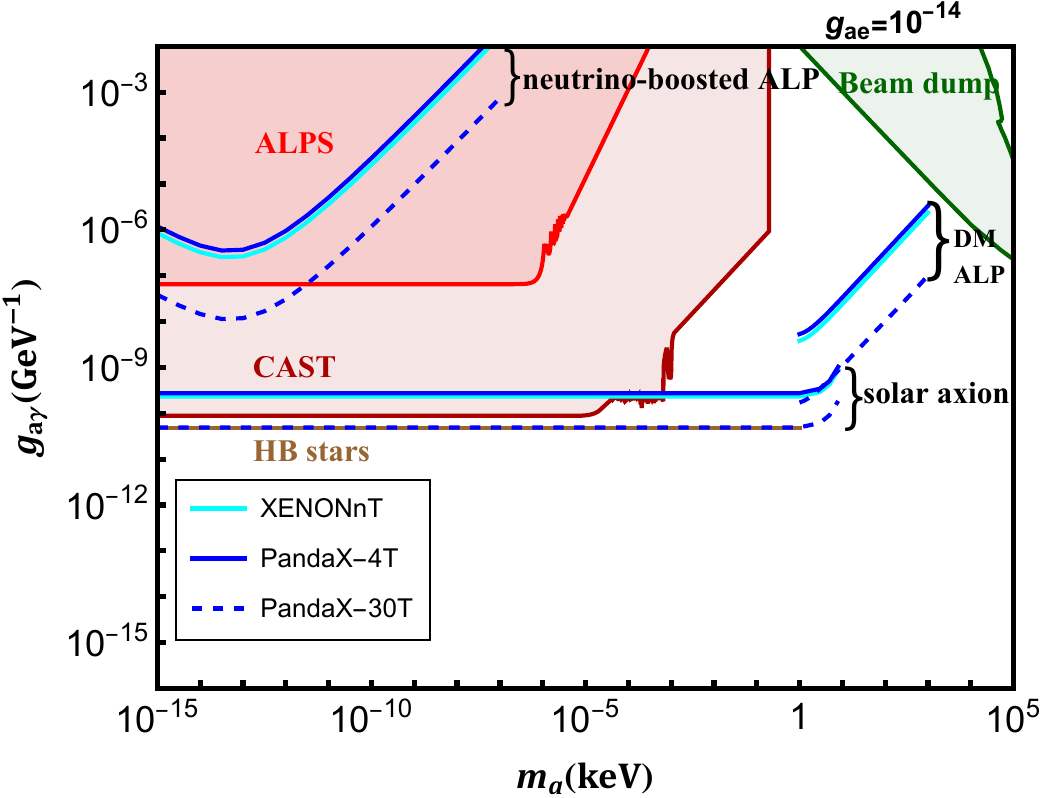}
		\end{minipage}%
		\begin{minipage}[c]{0.5\textwidth}
			\includegraphics[height=6cm,width=8cm]{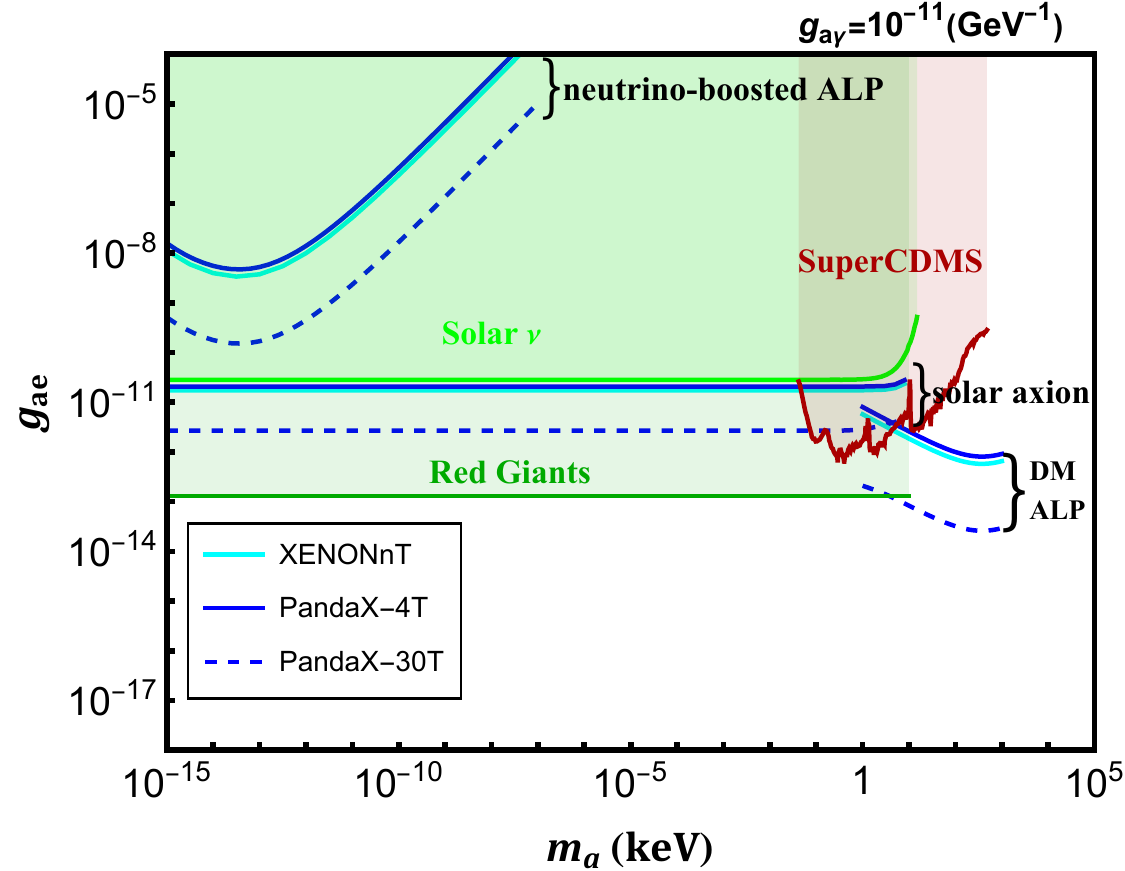}
		\end{minipage}
		\caption{$ 90\% $ C.L. limits on axion-photon or axion-electron coupling constants versus axion mass. Left-panel: we set $ g_{ae}=10^{-14} $. Right-panel: we fixed $ g_{a\gamma}=10^{-11}(\rm{GeV^{-1}}) $. } \label{fig4}
	\end{center}
\end{figure}

For the case $ g_{a\gamma}=0$ and $g_{ae}\neq0$, the number of events is given as $ N_{\rm{event}}=N_T \cdot \Phi_a \cdot \sigma^{\rm{IC}} $.  The $ 90\% $ C.L. limits are shown in the right-panel of the Fig.~\ref{fig3}, which shows $ g_{a e} $ versus $m_a$ for the XENONnT, PandaX-4T and PandaX-30T, respectively. 
Other limits shown in this plot includes astrophysical bounds from solar neutrino flux~(green)~\cite{Gondolo:2008dd} and red giants~(darker green)~\cite{Capozzi:2020cbu}, underground detectors SuperCDMS~(darker red)~\cite{SuperCDMS:2019jxx}.
The analysis is similar to the previous case, the difference is that the source of the solar axion flux is induced by the ABC processes. For DM axions,  $\sigma^{\rm{IC}} $ increases with $m_a$ for  $ m_a<1 \rm ~MeV $, so the upper limit on the $ g_{ae} $ inversely proportional to $m_a$, as shown in the plot. We did not choose a larger mass because ALP will decay in to electron-positron pair in this case. 

For the case $g_{a\gamma}\neq0$ and $g_{ae}\neq0$. We use the Eq.~(\ref{events}) directly to constrain the two couplings. 
We show in the left-panel of the Fig.~\ref{fig4}  the upper limits  on the $g_{a\gamma}$ by fixing  $ g_{ae}=10^{-14} $, from which we can see more stringent constraints put by the solar ALP flux. 
Since both $g_{ae}$ and $g_{a\gamma}$ are non-zero, the solar ALP flux is attributed to both the ABC processes and the Primakoff process. 

We further see that the limits of  the pandaX-30T in the future is almost comparable to the limits from astronomical observations, which is a very promising trend. 
Alternatively, we show  the exclusion limits on the $g_{ae}$ in the right-panel of the Fig.~\ref{fig4} by setting $ g_{a\gamma}=10^{-11}(\rm{GeV^{-1}}) $.
Since $g_{ae}$ is a free parameter and $g_{a\gamma}$ is small in this case, the solar ALP flux generated by the Primakoff effect is almost negligible compared to the ABC processes.

\begin{figure}[t]
	\begin{center}
		\begin{minipage}[c]{0.5\textwidth}
			\includegraphics[height=6cm,width=8cm]{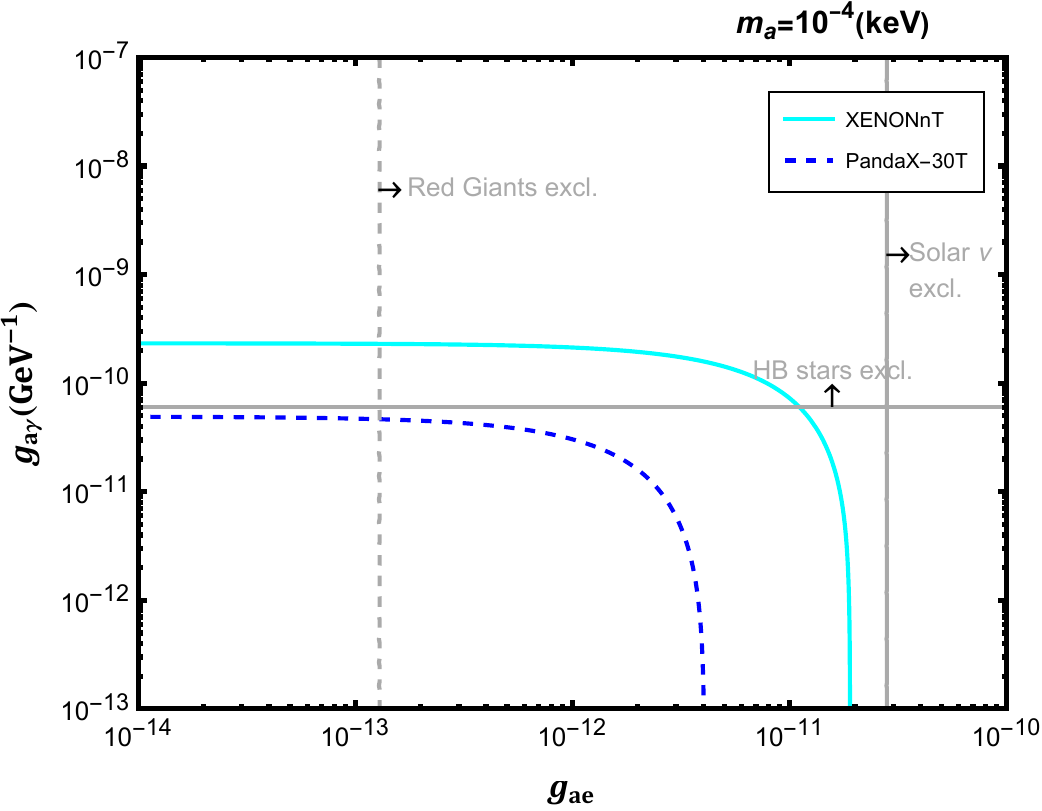}
		\end{minipage}%
		\begin{minipage}[c]{0.5\textwidth}
			\includegraphics[height=6cm,width=8cm]{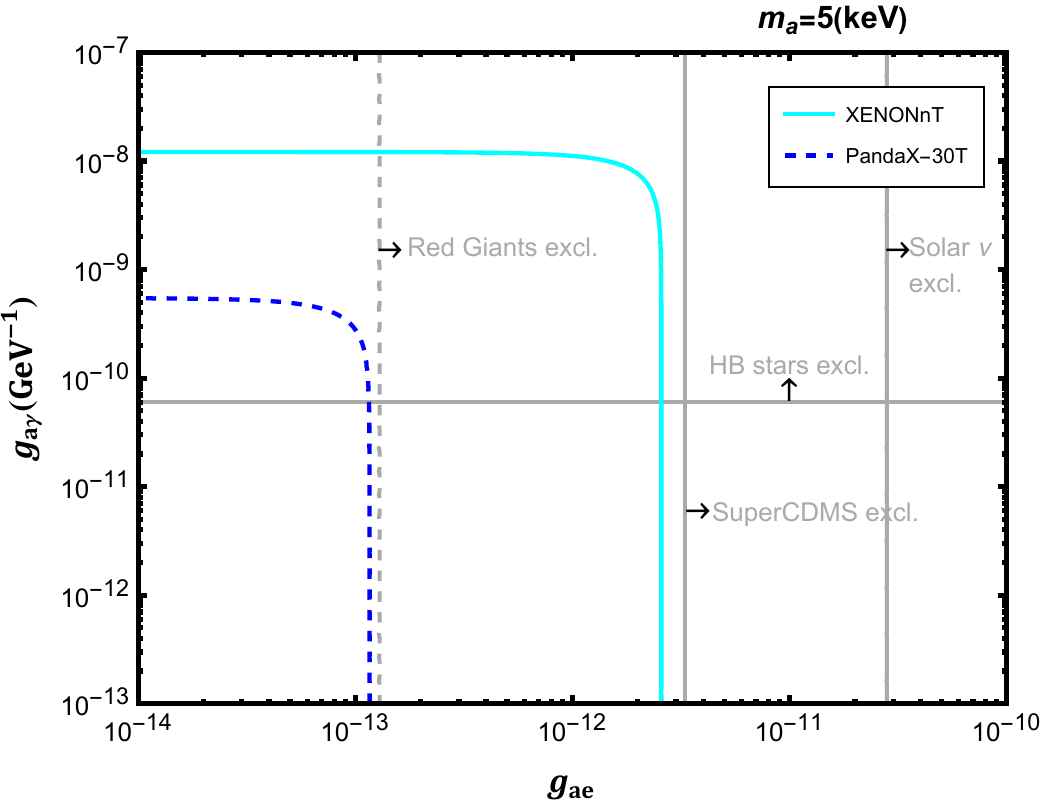}
		\end{minipage}
		\caption{$ 90\% $ C.L. limits on $ g_{ae}-g_{a\gamma} $ plane in XENONnT and PandaX-30T.  Gray lines repsresent the constraints from astrophysical observables including Red Giants~\cite{Capozzi:2020cbu} and the Horizontal Branch stars~\cite{Dolan:2022kul}, as well as the constraints from the solar neutrino~\cite{Gondolo:2008dd} and SuperCDMS~\cite{SuperCDMS:2019jxx}, with arrows denoting excluded regions.  We take $ m_a=10^{-4}~ \rm{keV}$ and $ m_a=5 ~\rm{keV}$ in the left-panel and right-panel, respectively. } \label{fig5}
	\end{center}
\end{figure}

Furthermore, we have scanned two parameters $ g_{a\gamma}$ and $ g_{ae} $ for different ALP masses. 
In Fig.~\ref{fig5}, we show the resulting $ 90\% $ C.L. constraints on the $ g_{ae}-g_{a\gamma} $ plane  given by the XENONnT and PandaX-30T. 
Since the constraint of  the pandaX-4T is very similar to that of the  XENONnT, we only show the results for XENONnT  for simplification.
By setting $ m_a=10^{-4}~ \rm{keV} $, the incoming ALP flux is mainly from the solar ALPs. 
When the ALP mass becomes larger, we can take $ m_a=5 ~\rm{keV} $, then the incoming axion flux is the from the DM ALP. 
Constraints for $ m_a=10^{-4} ~\rm{keV} $ and $ m_a=5~ \rm{keV} $ are listed in the left-panel and right-panel of the Fig.~\ref{fig5}, respectively.
The gray lines are exclusion limits arising from astrophysical observations. We find that even when adopting the most stringent constraints(Red Giants and Horizontal Branch stars), there is still  parameter space that survives in astrophysical observations but can be excluded by future projected direct detection experiments (see blue dashed lines in Fig.~\ref{fig5}), providing added confidence for further investigations.

\section{\label{sec: JUNO}JUNO Constraints}

The next-generation multi-purpose Jiangmen Underground Neutrino Observatory (JUNO)~\cite{JUNO:2015zny,JUNO:2015sjr,JUNO:2020hqc,JUNO:2020xtj,JUNO:2021vlw} is a state-of-the-art neutrino experiment primarily designed to investigate the properties of neutrinos, but it also has the potential to contribute to the search for new physics beyond the Standard Model, such as proton decay~\cite{JUNO:2015zny}, hidden sector particles~\cite{DEramo:2023buu} and axions~\cite{Lucente:2022esm}, etc. In this paper, we employ the JUNO detection experiment to set boundaries on the coupling constants of the axion. Thanks to its extensive exposure, we can significantly restrict the allowable values of the coupling constants.

One of the main components of the JUNO is the central detector, which is a massive, spherical Acrylic sphere filled with a high-purity liquid scintillator(LS). The detection medium is a linear alkylbenzene(LAB) composed of 19 carbon atoms $\rm C_{19}H_{32}$, which has excellent transparency, high flash point, low chemical reactivity and good light yield. The LS also doped with 3~g/L of 2,5-diphenyloxazole(PPO) and 15~mg/L of pbis-(o-methylstyryl)-benzene(bis-MSB). The density of the LS is 0.859g/ml, with a total of 20 ktons in a spherical container with a radius of 17.7 meters. The central detector is submerged in a cylindrical pool to protect it from the radioactivity of the surrounding rock. On the top of the water pool, muon tracker will be installed. We refer the reader to Refs.~\cite{JUNO:2015zny}  for more details about the JUNO.

\begin{figure}[t]
	\begin{center}
		\begin{minipage}[c]{0.5\textwidth}
			\includegraphics[height=6cm,width=8cm]{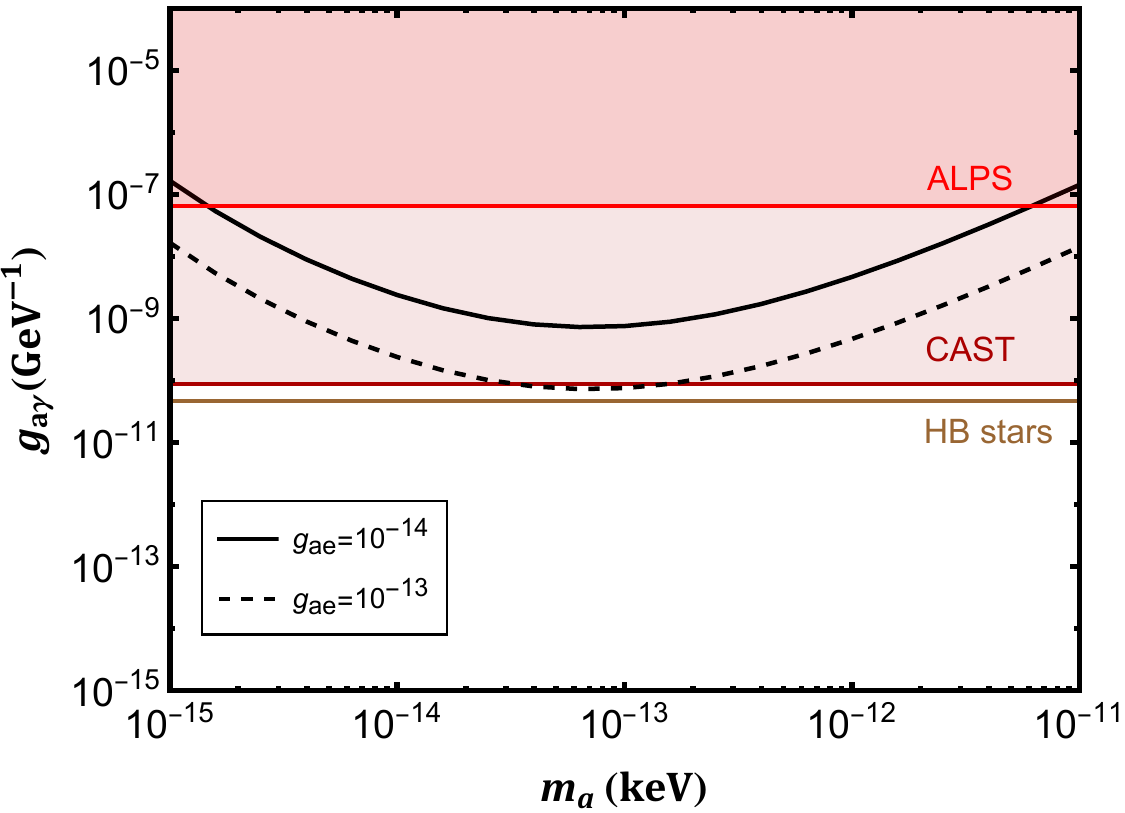}
		\end{minipage}%
		\begin{minipage}[c]{0.5\textwidth}
			\includegraphics[height=6cm,width=8cm]{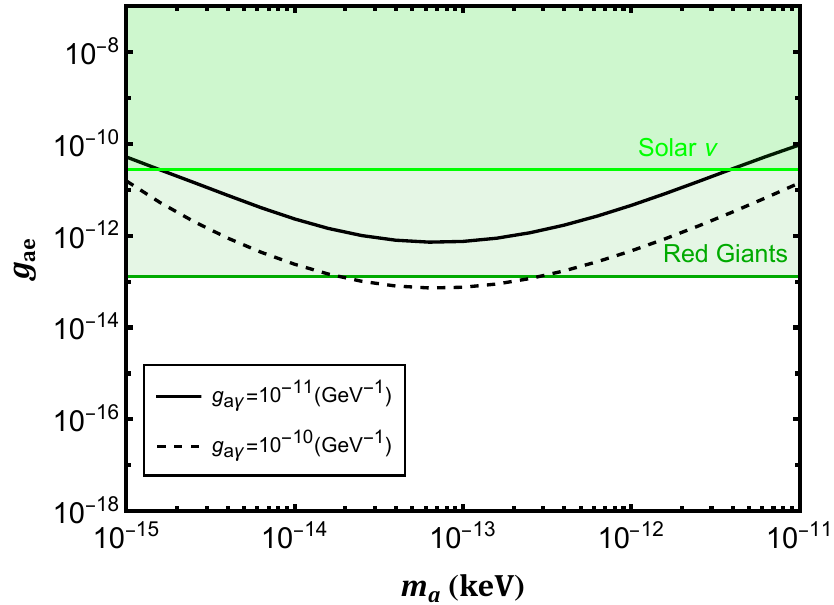}
		\end{minipage}
		\caption{$ 90\% $ C.L. limits on $ g_{a\gamma}~(g_{ae})-m_a $ planes in JUNO. Left-panel: Black solid line and black dashed line represent $g_{ae}=10^{-14}$ and $ g_{ae}=10^{-13}$, respectively. Right-panel: Black solid line and black dashed line represent $g_{a\gamma}=10^{-11}  \rm{ (GeV^{-1})}$ and $ g_{a\gamma}=10^{-10}  \rm{ (GeV^{-1})}$, respectively. } \label{fig6}
	\end{center}
\end{figure}

Due to the threshold of the detector, both the solar ALP flux and the DM ALP flux are almost undetectable in the JUNO, leaving the supernova neutrino boosted ALP as the only possible way out.
The axion fluxes have been given in Section \ref{sec: direct detection} Part A. 
Similar to the discussion in in the Section \ref{sec: direct detection} Part B, we still use $ \sigma^{\rm Tot}=\left(\sigma^{\rm{IP}}+\sigma^{\rm{IC}}+\sigma^{\rm{IT}}\right) $ to detect axions, and the expected number of events per unit time is
\begin{eqnarray}
		N_{\rm{event}}=N_e \cdot \Phi_a \cdot \left(\sigma^{\rm{IP}}+\sigma^{\rm{IC}}+\sigma^{\rm{IT}}\right) \cdot \mathcal{R} \cdot \epsilon \; ,
\end{eqnarray}
where $ N_e\simeq 5.5\times 10^{33} $ is the number of electrons  in the 16.2 kton fiducial volume. $ \mathcal{R} $ and $ \epsilon $ represent the detector energy resolution and the detector efficiency, respectively. 
The following discussion assumes that $ \epsilon=1 $ and over the energy threshold. 
Here, we use the likelihood function in Ref.~\cite{JUNO:2021vlw} to conduct numerical analysis. 
Ref.~\cite{DEramo:2023buu} shows the total number of events($ S $) expected to be detected over a ten-year period, $ S=97 $, corresponding to a $ 90\% $ C.L. sensitivity. Therefore, we use $ S_{\rm limit}=97 $ counts for 10 years data. The number of events used to constrain the upper limit on the axion coupling constants satisfy $ S=N_{\rm event}\cdot T \leq S_{\rm limit}$, and $ T=10 $ years. In this way, the numerical results we obtained are shown in the Fig.\ref{fig6}.

We choose to forecast the sensitivity at the $ 90\% $ C.L. to make a direct comparison with constraints of  DM direct detection experiments. Our result shows that the exclusion limits of the JUNO experiment are enhanced by approximately two orders of magnitude compared to direct detection experiments, owing to the substantial fiducial volume of JUNO. It is worth noting that even in the case of the next-generation PandaX-30T experiment, JUNO is likely to provide constraints on ALP couplings that are approximately one order of magnitude tighter. 
The most stringent constraints on coupling constants have been reached $g_{a \gamma}<7\times10^{-10} \rm{ GeV^{-1}}$ and $  g_{ae}<7.5 \times10^{-13} $, which are better than constraints from astrophysical observations. It should be mentioned that these are very conservative limits, as we have considered small values for both $ g_{ae} $(left-panel) and $g_{a \gamma}$(right-panel) in the Fig.\ref{fig6}.
\section{\label{sec: conclusion}Conclusion}

The direct detection of ALP is an important issue as ALP itself  serves a promising DM candidate and may solve the strong CP problem.  In addition to cavity experiments, which measure $g_{a\gamma}^{}$ via oscillations of ALP into photon in electromagnetic fields, direct detection experiments may also detect ALP via the inverse compton process or inverse Primakoff process. In this paper, we pay special attention to the calculation of the process $a+e\to e+ \gamma$, and show the interference between the inverse primakoff and inverse Compton amplitudes. We have shown constraints of projected direct detection experiments on couplings $g_{ae}$ and $g_{a\gamma}$. We further studied the constraints of the JUNO Cherenkov detector on these couplings.  This study provides theoretical support for the detections of ALP.

\begin{acknowledgments}
This work was supported by the National Natural Science Foundation of China (NSFC) (Grants No. 11775025 and No. 12175027).
\end{acknowledgments}

\bibliography{refs}

\begin{thebibliography}{73}
\expandafter\ifx\csname natexlab\endcsname\relax\def\natexlab#1{#1}\fi
\expandafter\ifx\csname bibnamefont\endcsname\relax
  \def\bibnamefont#1{#1}\fi
\expandafter\ifx\csname bibfnamefont\endcsname\relax
  \def\bibfnamefont#1{#1}\fi
\expandafter\ifx\csname citenamefont\endcsname\relax
  \def\citenamefont#1{#1}\fi
\expandafter\ifx\csname url\endcsname\relax
  \def\url#1{\texttt{#1}}\fi
\expandafter\ifx\csname urlprefix\endcsname\relax\def\urlprefix{URL }\fi
\providecommand{\bibinfo}[2]{#2}
\providecommand{\eprint}[2][]{\url{#2}}

\bibitem[{\citenamefont{Peccei and Quinn}(1977{\natexlab{a}})}]{Peccei:1977hh}
\bibinfo{author}{\bibfnamefont{R.~D.} \bibnamefont{Peccei}} \bibnamefont{and}
  \bibinfo{author}{\bibfnamefont{H.~R.} \bibnamefont{Quinn}},
  \bibinfo{journal}{Phys. Rev. Lett.} \textbf{\bibinfo{volume}{38}},
  \bibinfo{pages}{1440} (\bibinfo{year}{1977}{\natexlab{a}}).

\bibitem[{\citenamefont{Peccei and Quinn}(1977{\natexlab{b}})}]{Peccei:1977ur}
\bibinfo{author}{\bibfnamefont{R.~D.} \bibnamefont{Peccei}} \bibnamefont{and}
  \bibinfo{author}{\bibfnamefont{H.~R.} \bibnamefont{Quinn}},
  \bibinfo{journal}{Phys. Rev. D} \textbf{\bibinfo{volume}{16}},
  \bibinfo{pages}{1791} (\bibinfo{year}{1977}{\natexlab{b}}).

\bibitem[{\citenamefont{Weinberg}(1978)}]{Weinberg:1977ma}
\bibinfo{author}{\bibfnamefont{S.}~\bibnamefont{Weinberg}},
  \bibinfo{journal}{Phys. Rev. Lett.} \textbf{\bibinfo{volume}{40}},
  \bibinfo{pages}{223} (\bibinfo{year}{1978}).

\bibitem[{\citenamefont{Wilczek}(1978)}]{Wilczek:1977pj}
\bibinfo{author}{\bibfnamefont{F.}~\bibnamefont{Wilczek}},
  \bibinfo{journal}{Phys. Rev. Lett.} \textbf{\bibinfo{volume}{40}},
  \bibinfo{pages}{279} (\bibinfo{year}{1978}).

\bibitem[{\citenamefont{Kim and Carosi}(2010)}]{Kim:2008hd}
\bibinfo{author}{\bibfnamefont{J.~E.} \bibnamefont{Kim}} \bibnamefont{and}
  \bibinfo{author}{\bibfnamefont{G.}~\bibnamefont{Carosi}},
  \bibinfo{journal}{Rev. Mod. Phys.} \textbf{\bibinfo{volume}{82}},
  \bibinfo{pages}{557} (\bibinfo{year}{2010}), \bibinfo{note}{[Erratum:
  Rev.Mod.Phys. 91, 049902 (2019)]}, \eprint{0807.3125}.

\bibitem[{\citenamefont{Feng}(2010)}]{Feng:2010gw}
\bibinfo{author}{\bibfnamefont{J.~L.} \bibnamefont{Feng}},
  \bibinfo{journal}{Ann. Rev. Astron. Astrophys.}
  \textbf{\bibinfo{volume}{48}}, \bibinfo{pages}{495} (\bibinfo{year}{2010}),
  \eprint{1003.0904}.

\bibitem[{\citenamefont{Tam and Yang}(2011)}]{Tam:2011tz}
\bibinfo{author}{\bibfnamefont{H.}~\bibnamefont{Tam}} \bibnamefont{and}
  \bibinfo{author}{\bibfnamefont{Q.}~\bibnamefont{Yang}}
  (\bibinfo{year}{2011}), \eprint{1108.3362}.

\bibitem[{\citenamefont{Marsh}(2016)}]{Marsh:2015xka}
\bibinfo{author}{\bibfnamefont{D.~J.~E.} \bibnamefont{Marsh}},
  \bibinfo{journal}{Phys. Rept.} \textbf{\bibinfo{volume}{643}},
  \bibinfo{pages}{1} (\bibinfo{year}{2016}), \eprint{1510.07633}.

\bibitem[{\citenamefont{Spector}(2019)}]{Spector:2019ooq}
\bibinfo{author}{\bibfnamefont{A.}~\bibnamefont{Spector}}
  (\bibinfo{collaboration}{ALPS}), in \emph{\bibinfo{booktitle}{{14th Patras
  Workshop on Axions, WIMPs and WISPs}}} (\bibinfo{year}{2019}),
  \eprint{1906.09011}.

\bibitem[{\citenamefont{Anastassopoulos et~al.}(2017)}]{CAST:2017uph}
\bibinfo{author}{\bibfnamefont{V.}~\bibnamefont{Anastassopoulos}}
  \bibnamefont{et~al.} (\bibinfo{collaboration}{CAST}),
  \bibinfo{journal}{Nature Phys.} \textbf{\bibinfo{volume}{13}},
  \bibinfo{pages}{584} (\bibinfo{year}{2017}), \eprint{1705.02290}.

\bibitem[{\citenamefont{Akerib et~al.}(2017)}]{LUX:2017glr}
\bibinfo{author}{\bibfnamefont{D.~S.} \bibnamefont{Akerib}}
  \bibnamefont{et~al.} (\bibinfo{collaboration}{LUX}), \bibinfo{journal}{Phys.
  Rev. Lett.} \textbf{\bibinfo{volume}{118}}, \bibinfo{pages}{261301}
  (\bibinfo{year}{2017}), \eprint{1704.02297}.

\bibitem[{\citenamefont{Aprile et~al.}(2022)}]{XENON:2022ltv}
\bibinfo{author}{\bibfnamefont{E.}~\bibnamefont{Aprile}} \bibnamefont{et~al.}
  (\bibinfo{collaboration}{XENON}), \bibinfo{journal}{Phys. Rev. Lett.}
  \textbf{\bibinfo{volume}{129}}, \bibinfo{pages}{161805}
  (\bibinfo{year}{2022}), \eprint{2207.11330}.

\bibitem[{\citenamefont{Redondo}(2013)}]{Redondo:2013wwa}
\bibinfo{author}{\bibfnamefont{J.}~\bibnamefont{Redondo}},
  \bibinfo{journal}{JCAP} \textbf{\bibinfo{volume}{12}}, \bibinfo{pages}{008}
  (\bibinfo{year}{2013}), \eprint{1310.0823}.

\bibitem[{\citenamefont{Derbin et~al.}(2012)\citenamefont{Derbin, Drachnev,
  Kayunov, and Muratova}}]{Derbin:2012yk}
\bibinfo{author}{\bibfnamefont{A.~V.} \bibnamefont{Derbin}},
  \bibinfo{author}{\bibfnamefont{I.~S.} \bibnamefont{Drachnev}},
  \bibinfo{author}{\bibfnamefont{A.~S.} \bibnamefont{Kayunov}},
  \bibnamefont{and} \bibinfo{author}{\bibfnamefont{V.~N.}
  \bibnamefont{Muratova}}, \bibinfo{journal}{JETP Lett.}
  \textbf{\bibinfo{volume}{95}}, \bibinfo{pages}{339} (\bibinfo{year}{2012}),
  \eprint{1206.4142}.

\bibitem[{\citenamefont{Sikivie}(1983)}]{Sikivie:1983ip}
\bibinfo{author}{\bibfnamefont{P.}~\bibnamefont{Sikivie}},
  \bibinfo{journal}{Phys. Rev. Lett.} \textbf{\bibinfo{volume}{51}},
  \bibinfo{pages}{1415} (\bibinfo{year}{1983}), \bibinfo{note}{[Erratum:
  Phys.Rev.Lett. 52, 695 (1984)]}.

\bibitem[{\citenamefont{Andriamonje et~al.}(2007)}]{CAST:2007jps}
\bibinfo{author}{\bibfnamefont{S.}~\bibnamefont{Andriamonje}}
  \bibnamefont{et~al.} (\bibinfo{collaboration}{CAST}), \bibinfo{journal}{JCAP}
  \textbf{\bibinfo{volume}{04}}, \bibinfo{pages}{010} (\bibinfo{year}{2007}),
  \eprint{hep-ex/0702006}.

\bibitem[{\citenamefont{Dolan et~al.}(2022)\citenamefont{Dolan, Hiskens, and
  Volkas}}]{Dolan:2022kul}
\bibinfo{author}{\bibfnamefont{M.~J.} \bibnamefont{Dolan}},
  \bibinfo{author}{\bibfnamefont{F.~J.} \bibnamefont{Hiskens}},
  \bibnamefont{and} \bibinfo{author}{\bibfnamefont{R.~R.}
  \bibnamefont{Volkas}}, \bibinfo{journal}{JCAP} \textbf{\bibinfo{volume}{10}},
  \bibinfo{pages}{096} (\bibinfo{year}{2022}), \eprint{2207.03102}.

\bibitem[{\citenamefont{Fu et~al.}(2017)}]{PandaX:2017ock}
\bibinfo{author}{\bibfnamefont{C.}~\bibnamefont{Fu}} \bibnamefont{et~al.}
  (\bibinfo{collaboration}{PandaX}), \bibinfo{journal}{Phys. Rev. Lett.}
  \textbf{\bibinfo{volume}{119}}, \bibinfo{pages}{181806}
  (\bibinfo{year}{2017}), \eprint{1707.07921}.

\bibitem[{\citenamefont{Knapen et~al.}(2017)\citenamefont{Knapen, Lin, Lou, and
  Melia}}]{Knapen:2016moh}
\bibinfo{author}{\bibfnamefont{S.}~\bibnamefont{Knapen}},
  \bibinfo{author}{\bibfnamefont{T.}~\bibnamefont{Lin}},
  \bibinfo{author}{\bibfnamefont{H.~K.} \bibnamefont{Lou}}, \bibnamefont{and}
  \bibinfo{author}{\bibfnamefont{T.}~\bibnamefont{Melia}},
  \bibinfo{journal}{Phys. Rev. Lett.} \textbf{\bibinfo{volume}{118}},
  \bibinfo{pages}{171801} (\bibinfo{year}{2017}), \eprint{1607.06083}.

\bibitem[{\citenamefont{Aad et~al.}(2021)}]{ATLAS:2020hii}
\bibinfo{author}{\bibfnamefont{G.}~\bibnamefont{Aad}} \bibnamefont{et~al.}
  (\bibinfo{collaboration}{ATLAS}), \bibinfo{journal}{JHEP}
  \textbf{\bibinfo{volume}{03}}, \bibinfo{pages}{243} (\bibinfo{year}{2021}),
  \bibinfo{note}{[Erratum: JHEP 11, 050 (2021)]}, \eprint{2008.05355}.

\bibitem[{\citenamefont{Agnes et~al.}(2023)}]{DarkSide:2022knj}
\bibinfo{author}{\bibfnamefont{P.}~\bibnamefont{Agnes}} \bibnamefont{et~al.}
  (\bibinfo{collaboration}{DarkSide}), \bibinfo{journal}{Phys. Rev. Lett.}
  \textbf{\bibinfo{volume}{130}}, \bibinfo{pages}{101002}
  (\bibinfo{year}{2023}), \eprint{2207.11968}.

\bibitem[{\citenamefont{Arisaka et~al.}(2013)\citenamefont{Arisaka, Beltrame,
  Ghag, Kaidi, Lung, Lyashenko, Peccei, Smith, and Ye}}]{Arisaka:2012pb}
\bibinfo{author}{\bibfnamefont{K.}~\bibnamefont{Arisaka}},
  \bibinfo{author}{\bibfnamefont{P.}~\bibnamefont{Beltrame}},
  \bibinfo{author}{\bibfnamefont{C.}~\bibnamefont{Ghag}},
  \bibinfo{author}{\bibfnamefont{J.}~\bibnamefont{Kaidi}},
  \bibinfo{author}{\bibfnamefont{K.}~\bibnamefont{Lung}},
  \bibinfo{author}{\bibfnamefont{A.}~\bibnamefont{Lyashenko}},
  \bibinfo{author}{\bibfnamefont{R.~D.} \bibnamefont{Peccei}},
  \bibinfo{author}{\bibfnamefont{P.}~\bibnamefont{Smith}}, \bibnamefont{and}
  \bibinfo{author}{\bibfnamefont{K.}~\bibnamefont{Ye}},
  \bibinfo{journal}{Astropart. Phys.} \textbf{\bibinfo{volume}{44}},
  \bibinfo{pages}{59} (\bibinfo{year}{2013}), \eprint{1209.3810}.

\bibitem[{\citenamefont{Irastorza and Redondo}(2018)}]{Irastorza:2018dyq}
\bibinfo{author}{\bibfnamefont{I.~G.} \bibnamefont{Irastorza}}
  \bibnamefont{and} \bibinfo{author}{\bibfnamefont{J.}~\bibnamefont{Redondo}},
  \bibinfo{journal}{Prog. Part. Nucl. Phys.} \textbf{\bibinfo{volume}{102}},
  \bibinfo{pages}{89} (\bibinfo{year}{2018}), \eprint{1801.08127}.

\bibitem[{\citenamefont{Carenza and De~la Torre~Luque}(2023)}]{Carenza:2022som}
\bibinfo{author}{\bibfnamefont{P.}~\bibnamefont{Carenza}} \bibnamefont{and}
  \bibinfo{author}{\bibfnamefont{P.}~\bibnamefont{De~la Torre~Luque}},
  \bibinfo{journal}{Eur. Phys. J. C} \textbf{\bibinfo{volume}{83}},
  \bibinfo{pages}{110} (\bibinfo{year}{2023}), \eprint{2210.17206}.

\bibitem[{\citenamefont{Buchmuller and Hoogeveen}(1990)}]{Buchmuller:1989rb}
\bibinfo{author}{\bibfnamefont{W.}~\bibnamefont{Buchmuller}} \bibnamefont{and}
  \bibinfo{author}{\bibfnamefont{F.}~\bibnamefont{Hoogeveen}},
  \bibinfo{journal}{Phys. Lett. B} \textbf{\bibinfo{volume}{237}},
  \bibinfo{pages}{278} (\bibinfo{year}{1990}).

\bibitem[{\citenamefont{Creswick et~al.}(1998)\citenamefont{Creswick, Avignone,
  Farach, Collar, Gattone, Nussinov, and Zioutas}}]{Creswick:1997pg}
\bibinfo{author}{\bibfnamefont{R.~J.} \bibnamefont{Creswick}},
  \bibinfo{author}{\bibfnamefont{F.~T.} \bibnamefont{Avignone},
  \bibfnamefont{III}}, \bibinfo{author}{\bibfnamefont{H.~A.}
  \bibnamefont{Farach}}, \bibinfo{author}{\bibfnamefont{J.~I.}
  \bibnamefont{Collar}}, \bibinfo{author}{\bibfnamefont{A.~O.}
  \bibnamefont{Gattone}},
  \bibinfo{author}{\bibfnamefont{S.}~\bibnamefont{Nussinov}}, \bibnamefont{and}
  \bibinfo{author}{\bibfnamefont{K.}~\bibnamefont{Zioutas}},
  \bibinfo{journal}{Phys. Lett. B} \textbf{\bibinfo{volume}{427}},
  \bibinfo{pages}{235} (\bibinfo{year}{1998}), \eprint{hep-ph/9708210}.

\bibitem[{\citenamefont{Avignone et~al.}(1988)\citenamefont{Avignone, Baktash,
  Barker, Calaprice, Dunford, Haxton, Kahana, Kouzes, Miley, and
  Moltz}}]{Avignone:1988bv}
\bibinfo{author}{\bibfnamefont{F.~T.} \bibnamefont{Avignone}},
  \bibinfo{author}{\bibfnamefont{C.}~\bibnamefont{Baktash}},
  \bibinfo{author}{\bibfnamefont{W.~C.} \bibnamefont{Barker}},
  \bibinfo{author}{\bibfnamefont{F.~P.} \bibnamefont{Calaprice}},
  \bibinfo{author}{\bibfnamefont{R.~W.} \bibnamefont{Dunford}},
  \bibinfo{author}{\bibfnamefont{W.~C.} \bibnamefont{Haxton}},
  \bibinfo{author}{\bibfnamefont{D.}~\bibnamefont{Kahana}},
  \bibinfo{author}{\bibfnamefont{R.~T.} \bibnamefont{Kouzes}},
  \bibinfo{author}{\bibfnamefont{H.~S.} \bibnamefont{Miley}}, \bibnamefont{and}
  \bibinfo{author}{\bibfnamefont{D.~M.} \bibnamefont{Moltz}},
  \bibinfo{journal}{Phys. Rev. D} \textbf{\bibinfo{volume}{37}},
  \bibinfo{pages}{618} (\bibinfo{year}{1988}).

\bibitem[{\citenamefont{Chanda et~al.}(1988)\citenamefont{Chanda, Nieves, and
  Pal}}]{Chanda:1987ax}
\bibinfo{author}{\bibfnamefont{R.}~\bibnamefont{Chanda}},
  \bibinfo{author}{\bibfnamefont{J.~F.} \bibnamefont{Nieves}},
  \bibnamefont{and} \bibinfo{author}{\bibfnamefont{P.~B.} \bibnamefont{Pal}},
  \bibinfo{journal}{Phys. Rev. D} \textbf{\bibinfo{volume}{37}},
  \bibinfo{pages}{2714} (\bibinfo{year}{1988}).

\bibitem[{\citenamefont{Brodsky et~al.}(1986)\citenamefont{Brodsky, Mottola,
  Muzinich, and Soldate}}]{Brodsky:1986mi}
\bibinfo{author}{\bibfnamefont{S.~J.} \bibnamefont{Brodsky}},
  \bibinfo{author}{\bibfnamefont{E.}~\bibnamefont{Mottola}},
  \bibinfo{author}{\bibfnamefont{I.~J.} \bibnamefont{Muzinich}},
  \bibnamefont{and} \bibinfo{author}{\bibfnamefont{M.}~\bibnamefont{Soldate}},
  \bibinfo{journal}{Phys. Rev. Lett.} \textbf{\bibinfo{volume}{56}},
  \bibinfo{pages}{1763} (\bibinfo{year}{1986}), \bibinfo{note}{[Erratum:
  Phys.Rev.Lett. 57, 502 (1986)]}.

\bibitem[{\citenamefont{Li and Zhang}(2022)}]{Li:2022xqh}
\bibinfo{author}{\bibfnamefont{T.}~\bibnamefont{Li}} \bibnamefont{and}
  \bibinfo{author}{\bibfnamefont{R.-J.} \bibnamefont{Zhang}},
  \bibinfo{journal}{Phys. Rev. D} \textbf{\bibinfo{volume}{106}},
  \bibinfo{pages}{095034} (\bibinfo{year}{2022}), \eprint{2208.02696}.

\bibitem[{\citenamefont{Lucente et~al.}(2022)\citenamefont{Lucente, Nath,
  Capozzi, Giannotti, and Mirizzi}}]{Lucente:2022esm}
\bibinfo{author}{\bibfnamefont{G.}~\bibnamefont{Lucente}},
  \bibinfo{author}{\bibfnamefont{N.}~\bibnamefont{Nath}},
  \bibinfo{author}{\bibfnamefont{F.}~\bibnamefont{Capozzi}},
  \bibinfo{author}{\bibfnamefont{M.}~\bibnamefont{Giannotti}},
  \bibnamefont{and} \bibinfo{author}{\bibfnamefont{A.}~\bibnamefont{Mirizzi}},
  \bibinfo{journal}{Phys. Rev. D} \textbf{\bibinfo{volume}{106}},
  \bibinfo{pages}{123007} (\bibinfo{year}{2022}), \eprint{2209.11780}.

\bibitem[{\citenamefont{Dent et~al.}(2020{\natexlab{a}})\citenamefont{Dent,
  Dutta, Kim, Liao, Mahapatra, Sinha, and Thompson}}]{Dent:2019ueq}
\bibinfo{author}{\bibfnamefont{J.~B.} \bibnamefont{Dent}},
  \bibinfo{author}{\bibfnamefont{B.}~\bibnamefont{Dutta}},
  \bibinfo{author}{\bibfnamefont{D.}~\bibnamefont{Kim}},
  \bibinfo{author}{\bibfnamefont{S.}~\bibnamefont{Liao}},
  \bibinfo{author}{\bibfnamefont{R.}~\bibnamefont{Mahapatra}},
  \bibinfo{author}{\bibfnamefont{K.}~\bibnamefont{Sinha}}, \bibnamefont{and}
  \bibinfo{author}{\bibfnamefont{A.}~\bibnamefont{Thompson}},
  \bibinfo{journal}{Phys. Rev. Lett.} \textbf{\bibinfo{volume}{124}},
  \bibinfo{pages}{211804} (\bibinfo{year}{2020}{\natexlab{a}}),
  \eprint{1912.05733}.

\bibitem[{\citenamefont{Bhusal et~al.}(2021)\citenamefont{Bhusal, Houston, and
  Li}}]{Bhusal:2020bvx}
\bibinfo{author}{\bibfnamefont{A.}~\bibnamefont{Bhusal}},
  \bibinfo{author}{\bibfnamefont{N.}~\bibnamefont{Houston}}, \bibnamefont{and}
  \bibinfo{author}{\bibfnamefont{T.}~\bibnamefont{Li}}, \bibinfo{journal}{Phys.
  Rev. Lett.} \textbf{\bibinfo{volume}{126}}, \bibinfo{pages}{091601}
  (\bibinfo{year}{2021}), \eprint{2004.02733}.

\bibitem[{\citenamefont{Derevianko et~al.}(2010)\citenamefont{Derevianko,
  Dzuba, Flambaum, and Pospelov}}]{Derevianko:2010kz}
\bibinfo{author}{\bibfnamefont{A.}~\bibnamefont{Derevianko}},
  \bibinfo{author}{\bibfnamefont{V.~A.} \bibnamefont{Dzuba}},
  \bibinfo{author}{\bibfnamefont{V.~V.} \bibnamefont{Flambaum}},
  \bibnamefont{and} \bibinfo{author}{\bibfnamefont{M.}~\bibnamefont{Pospelov}},
  \bibinfo{journal}{Phys. Rev. D} \textbf{\bibinfo{volume}{82}},
  \bibinfo{pages}{065006} (\bibinfo{year}{2010}), \eprint{1007.1833}.

\bibitem[{\citenamefont{Donnelly et~al.}(1978)\citenamefont{Donnelly, Freedman,
  Lytel, Peccei, and Schwartz}}]{Donnelly:1978ty}
\bibinfo{author}{\bibfnamefont{T.~W.} \bibnamefont{Donnelly}},
  \bibinfo{author}{\bibfnamefont{S.~J.} \bibnamefont{Freedman}},
  \bibinfo{author}{\bibfnamefont{R.~S.} \bibnamefont{Lytel}},
  \bibinfo{author}{\bibfnamefont{R.~D.} \bibnamefont{Peccei}},
  \bibnamefont{and} \bibinfo{author}{\bibfnamefont{M.}~\bibnamefont{Schwartz}},
  \bibinfo{journal}{Phys. Rev. D} \textbf{\bibinfo{volume}{18}},
  \bibinfo{pages}{1607} (\bibinfo{year}{1978}).

\bibitem[{\citenamefont{Bellini et~al.}(2008)}]{Borexino:2008wiu}
\bibinfo{author}{\bibfnamefont{G.}~\bibnamefont{Bellini}} \bibnamefont{et~al.}
  (\bibinfo{collaboration}{Borexino}), \bibinfo{journal}{Eur. Phys. J. C}
  \textbf{\bibinfo{volume}{54}}, \bibinfo{pages}{61} (\bibinfo{year}{2008}).

\bibitem[{\citenamefont{Avignone et~al.}(1998)}]{SOLAX:1997lpz}
\bibinfo{author}{\bibfnamefont{F.~T.} \bibnamefont{Avignone},
  \bibfnamefont{III}} \bibnamefont{et~al.} (\bibinfo{collaboration}{SOLAX}),
  \bibinfo{journal}{Phys. Rev. Lett.} \textbf{\bibinfo{volume}{81}},
  \bibinfo{pages}{5068} (\bibinfo{year}{1998}), \eprint{astro-ph/9708008}.

\bibitem[{\citenamefont{Abe et~al.}(2021)\citenamefont{Abe, Hamaguchi, and
  Nagata}}]{Abe:2020mcs}
\bibinfo{author}{\bibfnamefont{T.}~\bibnamefont{Abe}},
  \bibinfo{author}{\bibfnamefont{K.}~\bibnamefont{Hamaguchi}},
  \bibnamefont{and} \bibinfo{author}{\bibfnamefont{N.}~\bibnamefont{Nagata}},
  \bibinfo{journal}{Phys. Lett. B} \textbf{\bibinfo{volume}{815}},
  \bibinfo{pages}{136174} (\bibinfo{year}{2021}), \eprint{2012.02508}.

\bibitem[{\citenamefont{Dent et~al.}(2020{\natexlab{b}})\citenamefont{Dent,
  Dutta, Newstead, and Thompson}}]{Dent:2020jhf}
\bibinfo{author}{\bibfnamefont{J.~B.} \bibnamefont{Dent}},
  \bibinfo{author}{\bibfnamefont{B.}~\bibnamefont{Dutta}},
  \bibinfo{author}{\bibfnamefont{J.~L.} \bibnamefont{Newstead}},
  \bibnamefont{and} \bibinfo{author}{\bibfnamefont{A.}~\bibnamefont{Thompson}},
  \bibinfo{journal}{Phys. Rev. Lett.} \textbf{\bibinfo{volume}{125}},
  \bibinfo{pages}{131805} (\bibinfo{year}{2020}{\natexlab{b}}),
  \eprint{2006.15118}.

\bibitem[{\citenamefont{Gao et~al.}(2020)\citenamefont{Gao, Liu, Wang, Wang,
  Xue, and Zhong}}]{Gao:2020wer}
\bibinfo{author}{\bibfnamefont{C.}~\bibnamefont{Gao}},
  \bibinfo{author}{\bibfnamefont{J.}~\bibnamefont{Liu}},
  \bibinfo{author}{\bibfnamefont{L.-T.} \bibnamefont{Wang}},
  \bibinfo{author}{\bibfnamefont{X.-P.} \bibnamefont{Wang}},
  \bibinfo{author}{\bibfnamefont{W.}~\bibnamefont{Xue}}, \bibnamefont{and}
  \bibinfo{author}{\bibfnamefont{Y.-M.} \bibnamefont{Zhong}},
  \bibinfo{journal}{Phys. Rev. Lett.} \textbf{\bibinfo{volume}{125}},
  \bibinfo{pages}{131806} (\bibinfo{year}{2020}), \eprint{2006.14598}.

\bibitem[{\citenamefont{Lin et~al.}(2023)\citenamefont{Lin, Wu, Wu, and
  Wong}}]{Lin:2022dbl}
\bibinfo{author}{\bibfnamefont{Y.-H.} \bibnamefont{Lin}},
  \bibinfo{author}{\bibfnamefont{W.-H.} \bibnamefont{Wu}},
  \bibinfo{author}{\bibfnamefont{M.-R.} \bibnamefont{Wu}}, \bibnamefont{and}
  \bibinfo{author}{\bibfnamefont{H.~T.-K.} \bibnamefont{Wong}},
  \bibinfo{journal}{Phys. Rev. Lett.} \textbf{\bibinfo{volume}{130}},
  \bibinfo{pages}{111002} (\bibinfo{year}{2023}), \eprint{2206.06864}.

\bibitem[{\citenamefont{Dimopoulos et~al.}(1986)\citenamefont{Dimopoulos,
  Starkman, and Lynn}}]{Dimopoulos:1986mi}
\bibinfo{author}{\bibfnamefont{S.}~\bibnamefont{Dimopoulos}},
  \bibinfo{author}{\bibfnamefont{G.~D.} \bibnamefont{Starkman}},
  \bibnamefont{and} \bibinfo{author}{\bibfnamefont{B.~W.} \bibnamefont{Lynn}},
  \bibinfo{journal}{Mod. Phys. Lett. A} \textbf{\bibinfo{volume}{1}},
  \bibinfo{pages}{491} (\bibinfo{year}{1986}).

\bibitem[{\citenamefont{Pospelov et~al.}(2008)\citenamefont{Pospelov, Ritz, and
  Voloshin}}]{Pospelov:2008jk}
\bibinfo{author}{\bibfnamefont{M.}~\bibnamefont{Pospelov}},
  \bibinfo{author}{\bibfnamefont{A.}~\bibnamefont{Ritz}}, \bibnamefont{and}
  \bibinfo{author}{\bibfnamefont{M.~B.} \bibnamefont{Voloshin}},
  \bibinfo{journal}{Phys. Rev. D} \textbf{\bibinfo{volume}{78}},
  \bibinfo{pages}{115012} (\bibinfo{year}{2008}), \eprint{0807.3279}.

\bibitem[{\citenamefont{Zhitnitsky and Skovpen}(1979)}]{Zhitnitsky:1979cn}
\bibinfo{author}{\bibfnamefont{A.~R.} \bibnamefont{Zhitnitsky}}
  \bibnamefont{and} \bibinfo{author}{\bibfnamefont{Y.~I.}
  \bibnamefont{Skovpen}}, \bibinfo{journal}{Sov. J. Nucl. Phys.}
  \textbf{\bibinfo{volume}{29}}, \bibinfo{pages}{513} (\bibinfo{year}{1979}).

\bibitem[{\citenamefont{Krauss et~al.}(1984)\citenamefont{Krauss, Moody, and
  Wilczek}}]{Krauss:1984gm}
\bibinfo{author}{\bibfnamefont{L.~M.} \bibnamefont{Krauss}},
  \bibinfo{author}{\bibfnamefont{J.~E.} \bibnamefont{Moody}}, \bibnamefont{and}
  \bibinfo{author}{\bibfnamefont{F.}~\bibnamefont{Wilczek}},
  \bibinfo{journal}{Phys. Lett. B} \textbf{\bibinfo{volume}{144}},
  \bibinfo{pages}{391} (\bibinfo{year}{1984}).

\bibitem[{\citenamefont{Mikaelian}(1978)}]{Mikaelian:1978jg}
\bibinfo{author}{\bibfnamefont{K.~O.} \bibnamefont{Mikaelian}},
  \bibinfo{journal}{Phys. Rev. D} \textbf{\bibinfo{volume}{18}},
  \bibinfo{pages}{3605} (\bibinfo{year}{1978}).

\bibitem[{\citenamefont{Fukugita
  et~al.}(1982{\natexlab{a}})\citenamefont{Fukugita, Watamura, and
  Yoshimura}}]{Fukugita:1982ep}
\bibinfo{author}{\bibfnamefont{M.}~\bibnamefont{Fukugita}},
  \bibinfo{author}{\bibfnamefont{S.}~\bibnamefont{Watamura}}, \bibnamefont{and}
  \bibinfo{author}{\bibfnamefont{M.}~\bibnamefont{Yoshimura}},
  \bibinfo{journal}{Phys. Rev. Lett.} \textbf{\bibinfo{volume}{48}},
  \bibinfo{pages}{1522} (\bibinfo{year}{1982}{\natexlab{a}}).

\bibitem[{\citenamefont{Fukugita
  et~al.}(1982{\natexlab{b}})\citenamefont{Fukugita, Watamura, and
  Yoshimura}}]{Fukugita:1982gn}
\bibinfo{author}{\bibfnamefont{M.}~\bibnamefont{Fukugita}},
  \bibinfo{author}{\bibfnamefont{S.}~\bibnamefont{Watamura}}, \bibnamefont{and}
  \bibinfo{author}{\bibfnamefont{M.}~\bibnamefont{Yoshimura}},
  \bibinfo{journal}{Phys. Rev. D} \textbf{\bibinfo{volume}{26}},
  \bibinfo{pages}{1840} (\bibinfo{year}{1982}{\natexlab{b}}).

\bibitem[{\citenamefont{Raffelt}(1986)}]{Raffelt:1985nk}
\bibinfo{author}{\bibfnamefont{G.~G.} \bibnamefont{Raffelt}},
  \bibinfo{journal}{Phys. Rev. D} \textbf{\bibinfo{volume}{33}},
  \bibinfo{pages}{897} (\bibinfo{year}{1986}).

\bibitem[{\citenamefont{Raffelt}(1988)}]{Raffelt:1987np}
\bibinfo{author}{\bibfnamefont{G.~G.} \bibnamefont{Raffelt}},
  \bibinfo{journal}{Phys. Rev. D} \textbf{\bibinfo{volume}{37}},
  \bibinfo{pages}{1356} (\bibinfo{year}{1988}).

\bibitem[{\citenamefont{Aghanim et~al.}(2020)}]{Planck:2018vyg}
\bibinfo{author}{\bibfnamefont{N.}~\bibnamefont{Aghanim}} \bibnamefont{et~al.}
  (\bibinfo{collaboration}{Planck}), \bibinfo{journal}{Astron. Astrophys.}
  \textbf{\bibinfo{volume}{641}}, \bibinfo{pages}{A6} (\bibinfo{year}{2020}),
  \bibinfo{note}{[Erratum: Astron.Astrophys. 652, C4 (2021)]},
  \eprint{1807.06209}.

\bibitem[{\citenamefont{Workman et~al.}(2022)}]{ParticleDataGroup:2022pth}
\bibinfo{author}{\bibfnamefont{R.~L.} \bibnamefont{Workman}}
  \bibnamefont{et~al.} (\bibinfo{collaboration}{Particle Data Group}),
  \bibinfo{journal}{PTEP} \textbf{\bibinfo{volume}{2022}},
  \bibinfo{pages}{083C01} (\bibinfo{year}{2022}).

\bibitem[{\citenamefont{Branca et~al.}(2017)}]{Branca:2016rez}
\bibinfo{author}{\bibfnamefont{A.}~\bibnamefont{Branca}} \bibnamefont{et~al.},
  \bibinfo{journal}{Phys. Rev. Lett.} \textbf{\bibinfo{volume}{118}},
  \bibinfo{pages}{021302} (\bibinfo{year}{2017}), \eprint{1607.07327}.

\bibitem[{\citenamefont{Lisanti}(2017)}]{Lisanti:2016jxe}
\bibinfo{author}{\bibfnamefont{M.}~\bibnamefont{Lisanti}}, in
  \emph{\bibinfo{booktitle}{{Theoretical Advanced Study Institute in Elementary
  Particle Physics}: {New Frontiers in Fields and Strings}}}
  (\bibinfo{year}{2017}), pp. \bibinfo{pages}{399--446}, \eprint{1603.03797}.

\bibitem[{\citenamefont{Vergados et~al.}(2022)\citenamefont{Vergados, Divari,
  and Ejiri}}]{Vergados:2021ejk}
\bibinfo{author}{\bibfnamefont{J.~D.} \bibnamefont{Vergados}},
  \bibinfo{author}{\bibfnamefont{P.~C.} \bibnamefont{Divari}},
  \bibnamefont{and} \bibinfo{author}{\bibfnamefont{H.}~\bibnamefont{Ejiri}},
  \bibinfo{journal}{Adv. High Energy Phys.} \textbf{\bibinfo{volume}{2022}},
  \bibinfo{pages}{7373365} (\bibinfo{year}{2022}), \eprint{2104.12213}.

\bibitem[{\citenamefont{Meng et~al.}(2021)}]{PandaX-4T:2021bab}
\bibinfo{author}{\bibfnamefont{Y.}~\bibnamefont{Meng}} \bibnamefont{et~al.}
  (\bibinfo{collaboration}{PandaX-4T}), \bibinfo{journal}{Phys. Rev. Lett.}
  \textbf{\bibinfo{volume}{127}}, \bibinfo{pages}{261802}
  (\bibinfo{year}{2021}), \eprint{2107.13438}.

\bibitem[{\citenamefont{Wang et~al.}(2023)\citenamefont{Wang, Lei, Ju, Liu,
  Zhou, Chen, Wang, Cui, Meng, and Zhao}}]{Wang:2023wrr}
\bibinfo{author}{\bibfnamefont{X.}~\bibnamefont{Wang}},
  \bibinfo{author}{\bibfnamefont{Z.}~\bibnamefont{Lei}},
  \bibinfo{author}{\bibfnamefont{Y.}~\bibnamefont{Ju}},
  \bibinfo{author}{\bibfnamefont{J.}~\bibnamefont{Liu}},
  \bibinfo{author}{\bibfnamefont{N.}~\bibnamefont{Zhou}},
  \bibinfo{author}{\bibfnamefont{Y.}~\bibnamefont{Chen}},
  \bibinfo{author}{\bibfnamefont{Z.}~\bibnamefont{Wang}},
  \bibinfo{author}{\bibfnamefont{X.}~\bibnamefont{Cui}},
  \bibinfo{author}{\bibfnamefont{Y.}~\bibnamefont{Meng}}, \bibnamefont{and}
  \bibinfo{author}{\bibfnamefont{L.}~\bibnamefont{Zhao}},
  \bibinfo{journal}{JINST} \textbf{\bibinfo{volume}{18}},
  \bibinfo{pages}{P05028} (\bibinfo{year}{2023}), \eprint{2301.06044}.

\bibitem[{\citenamefont{Liu et~al.}(2017)\citenamefont{Liu, Chen, and
  Ji}}]{Liu:2017drf}
\bibinfo{author}{\bibfnamefont{J.}~\bibnamefont{Liu}},
  \bibinfo{author}{\bibfnamefont{X.}~\bibnamefont{Chen}}, \bibnamefont{and}
  \bibinfo{author}{\bibfnamefont{X.}~\bibnamefont{Ji}},
  \bibinfo{journal}{Nature Phys.} \textbf{\bibinfo{volume}{13}},
  \bibinfo{pages}{212} (\bibinfo{year}{2017}), \eprint{1709.00688}.

\bibitem[{\citenamefont{Ehret et~al.}(2010)}]{Ehret:2010mh}
\bibinfo{author}{\bibfnamefont{K.}~\bibnamefont{Ehret}} \bibnamefont{et~al.},
  \bibinfo{journal}{Phys. Lett. B} \textbf{\bibinfo{volume}{689}},
  \bibinfo{pages}{149} (\bibinfo{year}{2010}), \eprint{1004.1313}.

\bibitem[{\citenamefont{Bergsma et~al.}(1985)}]{CHARM:1985anb}
\bibinfo{author}{\bibfnamefont{F.}~\bibnamefont{Bergsma}} \bibnamefont{et~al.}
  (\bibinfo{collaboration}{CHARM}), \bibinfo{journal}{Phys. Lett. B}
  \textbf{\bibinfo{volume}{157}}, \bibinfo{pages}{458} (\bibinfo{year}{1985}).

\bibitem[{\citenamefont{Riordan et~al.}(1987)}]{Riordan:1987aw}
\bibinfo{author}{\bibfnamefont{E.~M.} \bibnamefont{Riordan}}
  \bibnamefont{et~al.}, \bibinfo{journal}{Phys. Rev. Lett.}
  \textbf{\bibinfo{volume}{59}}, \bibinfo{pages}{755} (\bibinfo{year}{1987}).

\bibitem[{\citenamefont{Dolan et~al.}(2017)\citenamefont{Dolan, Ferber, Hearty,
  Kahlhoefer, and Schmidt-Hoberg}}]{Dolan:2017osp}
\bibinfo{author}{\bibfnamefont{M.~J.} \bibnamefont{Dolan}},
  \bibinfo{author}{\bibfnamefont{T.}~\bibnamefont{Ferber}},
  \bibinfo{author}{\bibfnamefont{C.}~\bibnamefont{Hearty}},
  \bibinfo{author}{\bibfnamefont{F.}~\bibnamefont{Kahlhoefer}},
  \bibnamefont{and}
  \bibinfo{author}{\bibfnamefont{K.}~\bibnamefont{Schmidt-Hoberg}},
  \bibinfo{journal}{JHEP} \textbf{\bibinfo{volume}{12}}, \bibinfo{pages}{094}
  (\bibinfo{year}{2017}), \bibinfo{note}{[Erratum: JHEP 03, 190 (2021)]},
  \eprint{1709.00009}.

\bibitem[{\citenamefont{Blumlein et~al.}(1991)}]{Blumlein:1990ay}
\bibinfo{author}{\bibfnamefont{J.}~\bibnamefont{Blumlein}}
  \bibnamefont{et~al.}, \bibinfo{journal}{Z. Phys. C}
  \textbf{\bibinfo{volume}{51}}, \bibinfo{pages}{341} (\bibinfo{year}{1991}).

\bibitem[{\citenamefont{Banerjee et~al.}(2020)}]{NA64:2020qwq}
\bibinfo{author}{\bibfnamefont{D.}~\bibnamefont{Banerjee}} \bibnamefont{et~al.}
  (\bibinfo{collaboration}{NA64}), \bibinfo{journal}{Phys. Rev. Lett.}
  \textbf{\bibinfo{volume}{125}}, \bibinfo{pages}{081801}
  (\bibinfo{year}{2020}), \eprint{2005.02710}.

\bibitem[{\citenamefont{Gondolo and Raffelt}(2009)}]{Gondolo:2008dd}
\bibinfo{author}{\bibfnamefont{P.}~\bibnamefont{Gondolo}} \bibnamefont{and}
  \bibinfo{author}{\bibfnamefont{G.~G.} \bibnamefont{Raffelt}},
  \bibinfo{journal}{Phys. Rev. D} \textbf{\bibinfo{volume}{79}},
  \bibinfo{pages}{107301} (\bibinfo{year}{2009}), \eprint{0807.2926}.

\bibitem[{\citenamefont{Capozzi and Raffelt}(2020)}]{Capozzi:2020cbu}
\bibinfo{author}{\bibfnamefont{F.}~\bibnamefont{Capozzi}} \bibnamefont{and}
  \bibinfo{author}{\bibfnamefont{G.}~\bibnamefont{Raffelt}},
  \bibinfo{journal}{Phys. Rev. D} \textbf{\bibinfo{volume}{102}},
  \bibinfo{pages}{083007} (\bibinfo{year}{2020}), \eprint{2007.03694}.

\bibitem[{\citenamefont{Aralis et~al.}(2020)}]{SuperCDMS:2019jxx}
\bibinfo{author}{\bibfnamefont{T.}~\bibnamefont{Aralis}} \bibnamefont{et~al.}
  (\bibinfo{collaboration}{SuperCDMS}), \bibinfo{journal}{Phys. Rev. D}
  \textbf{\bibinfo{volume}{101}}, \bibinfo{pages}{052008}
  (\bibinfo{year}{2020}), \bibinfo{note}{[Erratum: Phys.Rev.D 103, 039901
  (2021)]}, \eprint{1911.11905}.

\bibitem[{\citenamefont{An et~al.}(2016)}]{JUNO:2015zny}
\bibinfo{author}{\bibfnamefont{F.}~\bibnamefont{An}} \bibnamefont{et~al.}
  (\bibinfo{collaboration}{JUNO}), \bibinfo{journal}{J. Phys. G}
  \textbf{\bibinfo{volume}{43}}, \bibinfo{pages}{030401}
  (\bibinfo{year}{2016}), \eprint{1507.05613}.

\bibitem[{\citenamefont{Djurcic et~al.}(2015)}]{JUNO:2015sjr}
\bibinfo{author}{\bibfnamefont{Z.}~\bibnamefont{Djurcic}} \bibnamefont{et~al.}
  (\bibinfo{collaboration}{JUNO}) (\bibinfo{year}{2015}), \eprint{1508.07166}.

\bibitem[{\citenamefont{Abusleme et~al.}(2021{\natexlab{a}})}]{JUNO:2020hqc}
\bibinfo{author}{\bibfnamefont{A.}~\bibnamefont{Abusleme}} \bibnamefont{et~al.}
  (\bibinfo{collaboration}{JUNO}), \bibinfo{journal}{Chin. Phys. C}
  \textbf{\bibinfo{volume}{45}}, \bibinfo{pages}{023004}
  (\bibinfo{year}{2021}{\natexlab{a}}), \eprint{2006.11760}.

\bibitem[{\citenamefont{Abusleme et~al.}(2021{\natexlab{b}})}]{JUNO:2020xtj}
\bibinfo{author}{\bibfnamefont{A.}~\bibnamefont{Abusleme}} \bibnamefont{et~al.}
  (\bibinfo{collaboration}{JUNO}), \bibinfo{journal}{JHEP}
  \textbf{\bibinfo{volume}{03}}, \bibinfo{pages}{004}
  (\bibinfo{year}{2021}{\natexlab{b}}), \eprint{2011.06405}.

\bibitem[{\citenamefont{Abusleme et~al.}(2022)}]{JUNO:2021vlw}
\bibinfo{author}{\bibfnamefont{A.}~\bibnamefont{Abusleme}} \bibnamefont{et~al.}
  (\bibinfo{collaboration}{JUNO}), \bibinfo{journal}{Prog. Part. Nucl. Phys.}
  \textbf{\bibinfo{volume}{123}}, \bibinfo{pages}{103927}
  (\bibinfo{year}{2022}), \eprint{2104.02565}.

\bibitem[{\citenamefont{D'Eramo et~al.}(2023)\citenamefont{D'Eramo, Lucente,
  Nath, and Yun}}]{DEramo:2023buu}
\bibinfo{author}{\bibfnamefont{F.}~\bibnamefont{D'Eramo}},
  \bibinfo{author}{\bibfnamefont{G.}~\bibnamefont{Lucente}},
  \bibinfo{author}{\bibfnamefont{N.}~\bibnamefont{Nath}}, \bibnamefont{and}
  \bibinfo{author}{\bibfnamefont{S.}~\bibnamefont{Yun}} (\bibinfo{year}{2023}),
  \eprint{2305.14420}.

\end{thebibliography}

\end{document}